\documentclass[aps,pra,showpacs,twocolumn]{revtex4}
\usepackage{amsmath,amssymb,array}
\usepackage{graphicx}

\newcommand{\bvec}[1]{{\mathbf #1}}

\newcommand{\ket}[1]{\left| #1 \right>}

\newcommand{\beq}{\begin{eqnarray}}
\newcommand{\eeq}{\end{eqnarray}}

\begin{document}

\title
{Schwinger-Keldysh approach to out of equilibrium dynamics of the Bose Hubbard model
with time varying hopping}

\author{Malcolm P. Kennett$^1$ and Denis Dalidovich$^{1,2}$}

\affiliation
{
$^1$ Department of Physics, Simon Fraser University, 8888 University Drive, 
Burnaby, British Columbia V5A 1S6, Canada \\
$^2$ Perimeter Institute for Theoretical Physics, 31 Caroline Street North,
Waterloo, Ontario N2L 2Y5, Canada }

\date{\today}
\begin{abstract}
We study the real time dynamics of the Bose Hubbard model in the 
presence of time-dependent hopping allowing for a finite temperature initial state.  
We use the Schwinger-Keldysh
technique to find the real-time strong coupling action for the 
problem at both zero and finite temperature.  This action allows
for the description of both the superfluid and Mott insulating
phases.  We use this action 
to obtain dynamical equations for the superfluid order parameter as 
hopping is tuned in real time so that the system crosses the superfluid phase boundary.
We find that under a quench in the hopping, the system generically 
enters a metastable state in which the superfluid order parameter
has an oscillatory time dependence with a finite magnitude, but
disappears when averaged over a period.  We relate our results to recent
cold atom experiments.
\end{abstract}


\maketitle

\section{Introduction}

Ultracold atoms trapped in optical lattices 
\cite{BlochRev,Morsch,Bloch,Lewenstein} are highly versatile systems in 
which parameters can be tuned over wide ranges.  The ability to
tune these parameters in real time has opened the possibility
of studying the dynamic traversal of quantum phase 
transitions either in a ``quantum quench'' or with a 
more general time dependence.  This protocol has received
considerable interest 
\cite{Cardy,Mondal,Sengupta,Gritsev,PSSV,Sotiriadis,Sotiriadis2,Sciolla,BKL,Green}
as the resulting systems give examples of out of equilibrium dynamics in 
interacting quantum systems, a class of problem that is
still not fully understood.  

When bosons are cooled to lie in the lowest  Bloch band of
the periodic potential, their behaviour can be described using
the Bose-Hubbard model (BHM) \cite{Jaksch}.  The BHM
displays a transition between Mott-insulator and superfluid
phases as  the ratio of inter-site hopping $J$ to the on-site
repulsion $U$ is changed, as has been observed experimentally 
\cite{Greiner,Bloch,Gerbier3,Jimenez,Spielman}.
 This transition has been studied extensively theoretically 
and the equilibrium mean 
field solution is well known \cite{Fisher,Sheshadri,Sachdev,Herbut}.
More accurate determinations using quantum Monte Carlo
\cite{Scalettar,Krauth,Prokofev,Kato,Mahmud} 
and series expansions \cite{Monien}
verify the qualitative mean field picture \cite{Capogrosso}. 
In addition to cold atoms, there have also been proposals to realize
the BHM in photonic \cite{photon} and
polaritonic systems \cite{polariton}.

Experimentally there have been investigations of the 
transition from superfluid to Mott insulator or vice
versa by loading a condensate (or localized atoms)
into an optical lattice and then increasing or decreasing
the depth of the optical lattice \cite{Greiner,Bakr,Chen}.
Both the hopping between sites and the on-site interactions in the BHM
used to describe this situation depend on the strength of
the optical lattice potential \cite{Jaksch}, but the hopping is 
considerably more sensitive to the lattice depth 
than the interactions.  

Extensive theoretical effort has been expended on trying to 
understand the effects of time dependent $J/U$ in the BHM
(which can allow for a traversal of the phase transition).
Both sweeps from one phase to another, 
either gradually or as a quench \cite{Cucchietti,Clark,Lauchli,Kollath,Dziarmaga,Wang,Schutzhold,Fischer2,Altman,Polkovnikov,Zakrzewski,Cramer,PolkovnikovRev,Navez,Trefzger,Tiesinga,Lundh,Sau}
and periodic modulations with time \cite{Fischer1,Horiguchi,Fischer2,Gaul,Robertson,DasSarma,Poletti}
 similar to experiments in Refs.~\cite{Stoferle,Schori} have been considered.
A number of predictions have been made for these dynamics, including 
the time dependence of the decay of the superfluid order parameter for different
explicit forms of the time dependence of
$J(t)$ \cite{Schutzhold,Fischer2}; and of a wavevector dependent timescale for
freezing  \cite{Schutzhold,Lauchli,Fischer2} upon entering the Mott phase 
from the superfluid. Predictions for the
transition from the Mott phase to superfluid include the generation of 
vortices via the Kibble-Zurek mechanism, and scaling of time dependent
correlations with the quench timescale \cite{Cucchietti}. Such 
scaling (albeit with different exponents to those predicted in 
Ref.~\cite{Cucchietti}) was recently observed in experiments
by Chen {\it et al}. \cite{Chen}.  Studies of the extended BHM \cite{Schutzhold3}
and of quenches in the BHM \cite{Kollath,Fischer1,Sciolla}
suggest that non-equilibrium states can persist for considerable
times after a quench, especially for final states with small
values of $J/U$. In addition to the ratio $J/U$, time dependence of
other parameters, such as the chemical potential \cite{Lignier},
or even the lattice itself \cite{Hamma} have also been investigated.

The generation of out-of-equilibrium states from sweeps from 
the superfluid to the insulating phase (or vice versa) of the BHM 
is generic to dynamical traversals of quantum phase 
transitions 
\cite{Cardy,Mondal,Sengupta,Gritsev,PSSV,Sotiriadis,Sotiriadis2,Sciolla,BKL,Green}
and not limited to the BHM.  Experimentally it is not possible
to access zero temperature phase transitions, but as the effects
of such transitions extend to finite temperature, it is 
interesting to allow for thermal effects on the quench dynamics.
There has been considerable theoretical work on the BHM for non-zero
temperature \cite{Mahmud,Ho,Pollet,Gerbier,Stoof,Dickerscheid,Lu,Pupillo,Byczuk,Polak,Hoffmann,Hu,Trotzky}, but most has focused on the equilibrium 
properties of the model -- we allow for the effect of temperature in our 
out-of-equilibrium calculation by assuming a thermal initial state.

The approach we take to study the out of equilibrium 
dynamics of the BHM is to allow $J$ to be a function of time with $U$ constant.  
Our approach is sufficiently general
to allow for the inclusion of a trapping potential and time 
dependence in parameters other than $J$.
We construct a real-time effective action for the BHM using a 
strong coupling approach that can describe physics in both the 
superfluid and Mott insulating phases.  Various strong
coupling approaches have been proposed to allow description of both
phases in  equilibrium  \cite{Freericks,SenguptaDupuis,Freericks2,Trivedi,Tilahun},
and we generalize the imaginary time approach used in Ref.~\cite{SenguptaDupuis}
to real time by using the Schwinger-Keldysh formalism.  Several
authors have previously used 
Schwinger-Keldysh or closed time 
path \cite{Schwinger,Keldysh,Rammer,Semenoff,Landsman,Chou}  techniques
to study the Bose Hubbard model \cite{Rey1,Rey2,Calzetta,Gras,Grass2,Grassthesis,Temme}, but have
not focused on out-of-equilibrium dynamics.  

%Very recently there has been some
%work on the out of equilibrium dynamics of the BHM 
%with time dependent hopping carried out by Trefzger and 
%Sengupta \cite{Trefzger} using a projection operator technique. 

Given the assumption of time dependent hopping, we obtain the 
effective action within the Schwinger-Keldysh formalism.  We
then obtain the saddle point equations of motion, which we
are able to simplify to derive a mean field equation for the dynamics of
the superfluid order parameter during a quantum quench from the
superfluid phase to the insulating phase of the BHM at fixed
chemical potential.  We find that generically the solutions 
we obtain correspond to a final metastable
state in which the superfluid order parameter oscillates with a
finite magnitude, but averages to zero over a period of oscillation.
We note that the form of the metastable state depends on the 
value of the chemical potential and relate our results to work
showing that global mass redistribution is important
for the equilibration of cold atoms in traps after a quantum quench \cite{Natu}.

This paper is structured as follows.  In Sec.~\ref{sec:action} we
derive the effective action using the Schwinger-Keldysh/closed time path (CTP)
technique and in Sec.~\ref{sec:eqnmotion} we study the saddle point 
equations of motion for order parameter dynamics.  
In Sec.~\ref{sec:disc} we conclude and discuss our results.

\section{Effective action}
\label{sec:action}
In this section we discuss the application of the Schwinger-Keldysh 
technique to the Bose Hubbard model and derive a strong-coupling 
effective action for the model. 
The Hamiltonian for the Bose Hubbard model takes the form

\begin{eqnarray}
\hat{H}_{BH} & = & - \sum_{<ij>} J_{ij}\left(\hat{a}_i^\dagger \hat{a}_j
 + \hat{a}_j^\dagger \hat{a}_i\right) \nonumber \\
& &  + \frac{U}{2} \sum_j \hat{n}_j(\hat{n}_j -1) 
- \mu \sum_j \hat{n}_j , \nonumber \\
& = & \hat{H}_J + \hat{H}_0, \nonumber
\end{eqnarray}
where $\hat{a}_i$ and  $\hat{a}_i^\dagger$ are annihilation and creation 
operators for bosons on site $i$ respectively, $\hat{n}_i = 
\hat{a}^\dagger_i \hat{a}_i$ is the number operator, 
$U$ the interaction strength, and $\mu$ the chemical potential.
 The Hamiltonian
$$\hat{H}_0 = \hat{H}_U -\mu\hat{N} = \frac{U}{2} \sum_i \hat{n}_i (\hat{n}_i -1) - \mu \sum_i \hat{n}_i,$$ 
contains only single site terms, and $\hat{H}_J$ contains all of the 
hopping terms -- we allow for the possibility that the hopping amplitude 
$J_{ij}$ between sites $i$ and $j$ may be time dependent.

\subsection{Schwinger-Keldysh technique}
The Schwinger-Keldysh \cite{Schwinger,Keldysh} or closed time path (CTP)
technique \cite{Rammer,Semenoff,Landsman,Chou} is an approach that allows
a description of out of equilibrium or equilibrium quantum phenomena 
within the same formalism. 
The usual approach to finite temperature calculations is to use the
Matsubara formalism, which is restricted to equilibrium, and requires
analytic continuation to obtain real time dynamics.  The advantage of
CTP methods is that the problem is formulated in real time so that 
out of equilibrium problems can be tackled and no analytic continuation 
is required -- the price to pay is that the number of fields in the 
theory doubles, a second copy of each field propagates backwards in 
time.  As discussed by e.g. Niemi and Semenoff \cite{Semenoff}, the 
notion of time ordering needs to be replaced by that of contour ordering 
in order to calculate Green's functions.

\begin{figure}[htb]
\includegraphics[width=4cm,angle=270]{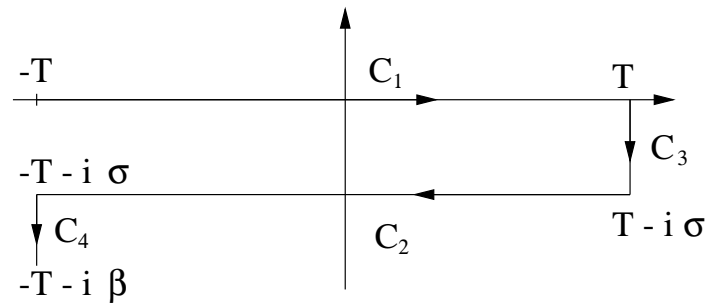}
\caption{Contour for the Schwinger-Keldysh technique for a system with 
inverse temperature $\beta$.  The value of $\sigma$ is arbitrary in 
the interval $[0,\beta]$ (Ref.~\cite{Landsman}). }
\label{fig:thermalcontour}
\end{figure}
For a thermal initial state, as we will assume here, the generating 
functional ${\mathcal Z}$ factorizes \cite{Semenoff}:
$$ {\mathcal Z} = {\mathcal Z}_{C_1 \cup C_2} {\mathcal Z}_{C_3 \cup C_4},$$
with $C_1$, $C_2$, $C_3$, and $C_4$ contour segments as illustrated in
Fig.~\ref{fig:thermalcontour}.  The
value of $0 \leq \sigma \leq \beta$ is arbitrary \cite{Landsman} 
-- we work with $\sigma = 0$ for simplicity.  

\subsection{Effective action for the Bose Hubbard model}
We may write a path integral for the 
generating functional of the BHM:

\begin{eqnarray}
{\mathcal Z} = \int [{\mathcal D}a^*][{\mathcal D}a] e^{iS_{\rm BHM}[a^*,a] 
},
\label{eq:genfunc}
\end{eqnarray}
where $a$ is a bosonic field and 
we omit source fields and set $\hbar = 1$. 
The action for the Bose Hubbard model has the form

\begin{eqnarray}
S_{\rm BHM} = \int_{-\infty}^\infty dt \, \left[a_{ia}^*(t) \left(i\partial_t\right)\tau^3_{ab} a_{ib}(t)\right]
 + S_J + S_U ,
\end{eqnarray}
where
\begin{eqnarray}
S_J & = & \int_{-\infty}^\infty dt \sum_{<ij>} J_{ij} \left[a_{ia}^*(t) \tau^3_{ab} a_{jb}(t) + a_{ja}^*(t) \tau^3_{ab} a_{ib}(t)
\right] , \nonumber \\ & & 
\end{eqnarray}
and $S_U$ is the action associated with $H_0$, where $a_{ia}$ 
is the field at site $i$ on contour $a$, where $a = 1$ or 2.
We use notation such that $\tau^i$ is the $i^{\rm th}$ Pauli matrix, 
acting in Keldysh space rather than spin space. 

We perform a Keldysh rotation so that 
$$ \left(\begin{array}{c} a_1(t) \\ a_2(t) \end{array}\right) \longrightarrow \left(\begin{array}{c}\tilde{a}_q(t) \\ \tilde{a}_c(t) 
\end{array} \right) = \hat{L} \left(\begin{array}{c} a_1(t) \\ a_2(t) \end{array} \right),$$
where $a_q$ and $a_c$ are the quantum and classical components of the field
 respectively \cite{Weert,Cugliandolo,KCY,Grassthesis}, and
$$\hat{L} = \frac{1}{\sqrt{2}} \left(\begin{array}{cc} 1 & -1 \\ 1 & 1 \end{array} \right).$$ 
The effect of this on the action is that $\tau^3$ in the $1$, $2$ basis becomes $\tau^1$ in the $q$, $c$ 
basis, hence (dropping tildes)
$$ S_J = \sum_{<ij>} \int_{-\infty}^\infty J_{ij} \left[a_{ia}^*(t) \tau^1_{ab} a_{jb}(t) + a_{ja}^*(t) \tau^{1}_{ab} a_{ib}(t)
\right].$$

Unlike previous studies of the BHM using closed
time path techniques \cite{Rey1,Rey2,Calzetta,Gras,Grass2,Grassthesis,Temme},
we are interested in the problem in which the hopping varies as a 
function of time to cross from the superfluid to the Mott Insulating 
phase.  Hence we require a formalism that allows for an adequate 
description of both phases.  We thus generalize to real time the 
strong coupling method used in imaginary time by Sengupta and Dupuis 
\cite{SenguptaDupuis}.  The advantage of this approach, as pointed out 
in Ref.~\cite{SenguptaDupuis} is that it leads to a normalized 
spectral function, which allows for the calculation of the excitation 
spectrum and momentum distribution in the superfluid phase, whilst also
giving a good description of the Mott insulating phase.  A similar
equilibrium effective action based on the Keldysh approach was 
recently obtained in Refs.~\cite{Gras,Grass2,Grassthesis}.

The approach requires two Hubbard-Stratonovich transformations.
The first of these decouples the hopping term.  We 
introduce a Hubbard-Stratonovich field $\psi$ and make use of the identity
(derived in Appendix~\ref{app:HS})
\begin{eqnarray}
e^{-i(\xi^* \eta + \xi \eta^*)}
& = & \int \overline{\mathcal D}(\varphi_1,\varphi_1^*)
{\mathcal D}(\varphi_2,\varphi_2^*) e^{i(\varphi_2^* \varphi_1 +
\varphi_1^* \varphi_2)} \nonumber \\
& & \hspace*{0.5cm} \times 
e^{i(\varphi_1^* \xi + \varphi_1 \xi^* +\varphi_2^*\eta + \varphi_2 \eta^*)} ,
\end{eqnarray}
to write

\begin{eqnarray}
{\mathcal Z} = \int [{\mathcal D} \psi^*][{\mathcal D}\psi] e^{-\frac{i}{2}\int_{-\infty}^\infty dt \sum_{ij} 
\psi_{ia}^*(t)\tau^1_{ab}J_{ij}^{-1}\psi_{jb}(t)} e^{iW[\psi^*,\psi]}, \nonumber \\
\end{eqnarray}
with
$$ e^{iW[\psi^*,\psi]} = \left<e^{-i\int dt \sum_i \psi_{ia}^*(t) \tau^1_{ab} a_{ib}(t) + \psi_{ia}(t) \tau^1_{ab} a_{ib}^*(t)}
\right>_0,$$
where the average $\left<\ldots\right>_0$ is taken with respect to 
$$S_0 = \int_{-\infty}^\infty dt \sum_i \left[ a^*_{ia}(t) \left(i\partial_t\right) 
\tau^1_{ab} a_{ib}(t) \right] + S_U.$$

$W[\psi^*,\psi]$ can be used to calculate the $2n$ point 
connected Green's functions $G^{nc}$ for the bosonic field $a$ via:

\begin{widetext}

\begin{eqnarray}
G^{nc}_{ia_1\ldots a_n a_1^\prime \ldots a_n^\prime}(t_1, \ldots,t_n,t_1^\prime,\ldots,t_n^\prime) 
& = & 
 e^{-iW[0]} 
 \left. \left\{ \frac{(-1)^n \delta^{(2n)}\left[e^{iW[\psi^*,\psi]}\right]}{
\delta \psi^*_{ia_1}(t_1) \ldots \delta \psi^*_{ia_n}(t_n)
\delta \psi_{ia_n^\prime}(t_n^\prime) \delta \psi_{ia_1^\prime}(t_1^\prime)}
\right\} \right|_{\psi^* = \psi = 0} \nonumber \\
& = & i
 \left. \left\{ \frac{(-1)^n \delta^{(2n)}W[\psi^*,\psi]}{
\delta \psi^*_{ia_1}(t_1) \ldots \delta \psi^*_{ia_n}(t_n)
\delta \psi_{ia_n^\prime}(t_n^\prime) \delta \psi_{ia_1^\prime}(t_1^\prime)}
\right\} \right|_{\psi^* = \psi = 0} \nonumber \\
& = & i(-1)^n \tau^1_{a_1 b_1} \ldots \tau^1_{a_n b_n} \tau^1_{a_1^\prime b_1^\prime} \ldots 
\tau^1_{a_n^\prime b_n^\prime} \left<a_{ib_1}(t_1) \ldots a_{ib_n}(t_n) a^*_{ib_n^\prime}(t_n^\prime)
\ldots a^*_{ib_1^\prime}(t_1^\prime) \right>_0^c , \nonumber \\ & & 
\label{eq:Gcondef}
\end{eqnarray}
where the superscript $c$ indicates a connected function.
Note that the connected Green's function vanishes if not all sites are identical.
Thus, we may write (similarly to Ref.~\cite{SenguptaDupuis}): 
\begin{eqnarray}
iW[\psi^*,\psi] & = & i \sum_i \sum_{n=1}^\infty \frac{(-1)^n}{(n!)^2} 
\int_{-\infty}^\infty \left[\prod_{l=1}^n dt_l dt_l^\prime \right] 
\psi_{ia_1}^*(t_1) \ldots \psi_{ia_n}^*(t_n) 
\psi_{ia_n^\prime}(t_n^\prime) \ldots \psi_{ia_1^\prime} \nonumber \\
& & \times \tau^1_{a_1 b_1} \ldots \tau^1_{a_n b_n} 
\tau^1_{a_1^\prime b_1^\prime} \ldots \tau^1_{a_n^\prime b_n^\prime} 
G^{nc}_{i,b_1 \ldots b_n b_1^\prime \ldots b_n^\prime}(t_1, \ldots,t_n; t_1^\prime, \ldots, t_n^\prime) ,
\end{eqnarray}
and so
$$ e^{iW[\psi^*,\psi]} = e^{i\sum_{n=1}^\infty S^n_{\rm int}[\psi^*,\psi]},$$
where
\begin{eqnarray}
S^n_{\rm int} & = & 
\frac{(-1)^n}{(n!)^2} \sum_i 
\int_{-\infty}^\infty  \left[\prod_{l=1}^n dt_l dt_l^\prime \right]
 \psi_{ia_1}^*(t_1) \ldots \psi_{ia_n}^*(t_n) 
\psi_{ia_n^\prime}(t_n^\prime) \ldots \psi_{ia_1^\prime}(t_1^\prime) \nonumber \\
& & \times \tau^1_{a_1 b_1} \ldots \tau^1_{a_n b_n} 
\tau^1_{a_1^\prime b_1^\prime} \ldots \tau^1_{a_n^\prime b_n^\prime} 
G^{nc}_{i,b_1 \ldots b_n b_1^\prime \ldots b_n^\prime }(t_1, \ldots,t_n; t_1^\prime, \ldots, t_n^\prime)  .
\end{eqnarray}

Summarizing the effective action to quartic order after the first Hubbard-Stratonovich
transformation gives:

\begin{eqnarray}
S^I_{\rm eff}[\psi^*,\psi] & = & -\frac{1}{2} \int dt \sum_{ij} \psi_{ia}^*(t) (J_{ij})^{-1}
\tau^1_{ab} \psi_{ib}(t) 
- \int dt_1 dt_2 \sum_i \psi_{ia_1}^*(t_1) \tau^1_{a_1 b_1} G_{ib_1 b_2}(t_1,t_2)
\tau^1_{b_2 a_2} \psi_{ia_2}(t_2) \nonumber \\
& & + \frac{1}{4} \int dt_1 dt_2 dt_3 dt_4 \sum_i \psi_{ia_1}^*(t_1) \psi_{ia_2}^*(t_2)
\tau^1_{a_1 b_1} \tau^1_{a_2 b_2} G^{2c}_{i b_1 b_2 b_3 b_4}(t_1,t_2,t_3,t_4) 
\tau^1_{a_3 b_3} \tau^1_{a_4 b_4} \psi_{i a_3}(t_3) \psi_{i a_4}(t_4) . \nonumber \\ & &
\label{eq:firstHS}
\end{eqnarray}

\end{widetext}

We discuss how the mean field phase boundary at zero and finite temperature
may be obtained from Eq.~(\ref{eq:firstHS}) in Appendix~\ref{app:aside}.
Sengupta and Dupuis \cite{SenguptaDupuis} observed that although
the equilibrium action of the form obtained in Eq.~(\ref{eq:firstHS}) leads
to the correct mean field phase boundary, it leads
to an unphysical excitation spectrum in the superfluid phase.
This can be rectified by performing a second Hubbard-Stratonovich transformation
\cite{SenguptaDupuis}. Starting from
\begin{eqnarray}
{\mathcal Z} = \int [{\mathcal D} \psi^*][{\mathcal D}\psi] e^{-\frac{i}{2}\int_{-\infty}^\infty dt \sum_{ij}
\psi_{ia}^*(t)\tau^1_{ab}J_{ij}^{-1}\psi_{jb}(t)} e^{iW[\psi^*,\psi]} , \nonumber \\
\end{eqnarray}
introduce a field $z$ such that 
\begin{widetext}
\begin{eqnarray}
e^{-\frac{i}{2}\int_{-\infty}^\infty dt \sum_{ij}
\psi_{ia}^*(t)\tau^1_{ab}J_{ij}^{-1}\psi_{jb}(t)}
 = \int [{\mathcal D}z^*] [{\mathcal D}z] e^{i\int dt \sum_{ij} 
(2J_{ij})z^*_{ia}(t)\tau^1_{ab} z_{jb}(t) }
e^{i\int dt \sum_i \left[ z_{ia}^*(t) \tau^1_{ab} \psi_{ib}(t) + \psi_{ia}^*(t)
\tau^1_{ab} z_{ib}(t) \right]} ,
\end{eqnarray}
so we have
\begin{eqnarray}
{\mathcal Z}  = \int [{\mathcal D}z^*] [{\mathcal D}z] e^{i\int dt \sum_{ij}
(2J_{ij})z^*_{ia}(t)\tau^1_{ab} z_{jb}(t) }
 \int [{\mathcal D} \psi^*][{\mathcal D}\psi]
e^{i\int dt \sum_i \left[ z_{ia}^*(t) \tau^1_{ab} \psi_{ib}(t) + \psi_{ia}^*(t)
\tau^1_{ab} z_{ib}(t) \right]}
e^{iW[\psi^*,\psi]} .
\end{eqnarray}
As discussed earlier,
$$ e^{iW[\psi^*,\psi]} = e^{i\sum_{n=1}^\infty S^n_{\rm int}(\psi^*,\psi)}
 = e^{iS_G + i\sum_{n=2}^\infty S^n_{\rm int}(\psi^*,\psi)},$$
where $S_G$ is the quadratic term
$$S_G = -\sum_i \int dt_1 dt_2 \psi^*_{ia_1}(t_1) \tau^1_{a_1 b_1} G_{i b_1 b_2}(t_1,t_2)
\tau^1_{b_2 a_2} \psi_{i a_2}(t_2) ,$$
and let
\begin{eqnarray}
e^{i\tilde{W}(z^*,z)} & = & \int [{\mathcal D} \psi^*][{\mathcal D}\psi]
e^{iS_G + i\int dt \sum_i \left[z_{ia}^*(t) \tau^1_{ab} \psi_{ib}(t) + \psi_{ia}^*(t)
\tau^1_{ab} z_{ib}(t) \right]} 
e^{i\sum_{n=2}^\infty S^n_{\rm int}(\psi^*,\psi)} \nonumber \\
&  = & \left< e^{ i\int dt \sum_i \left[z_{ia}^*(t) \tau^1_{ab} \psi_{ib}(t) + \psi_{ia}^*(t)
\tau^1_{ab} z_{ib}(t) \right] + i\sum_{n=2}^\infty S^n_{\rm int}(\psi^*,\psi)}
\right>_{S_G} .
\end{eqnarray}
We next perform a cumulant expansion for $\tilde{W}(z^*,z)$ and keep only terms in the 
action that are not ``anomalous''  
(for further discussion see Refs.~\cite{SenguptaDupuis,Dupuis}) to obtain
$$ Z  = \int [{\mathcal D}z^*] [{\mathcal D}z] e^{iS_{\rm eff}^{II}[z^*,z]} ,$$
where in calculating the  effective action to quartic order in $z$, we  
truncated
$i\sum_{n=2}^\infty S^n_{\rm int} \to i S^2_{\rm int},$
with
$$  S^2_{\rm int} = \frac{1}{(2!)^2} \sum_i \int dt_1 dt_2 dt_1^\prime dt_2^\prime
\psi^*_{ia_1}(t_1) \psi^*_{ia_2}(t_2) \tau^1_{a_1 b_1} \tau^1_{a_2 b_2}
G^{2c}_{i b_1 b_2 b_2^\prime b_1^\prime}(t_1,t_2,t_1^\prime,t_2^\prime)
\tau^1_{b_2^\prime a_2^\prime} \tau^1_{b_1^\prime a_1^\prime} 
\psi_{ia_2^\prime}(t_2^\prime) \psi_{ia_1^\prime}(t_1^\prime) . $$
The effective action to quartic order in the $z$ fields is

\begin{eqnarray}
S_{\rm eff}^{II}[z^*,z] & = & \int dt \sum_{ij} z_{ia}^*(t) (2J_{ij}) \tau^1_{ab} z_{jb}(t)
 + \int dt_1 dt_2 \sum_i z^*_{i a_1}(t_1) \left[G_{ia_2 a_1}(t_2,t_1)\right]^{-1} z_{ia_2}(t_2) \nonumber \\& &
 + \frac{1}{4} \int dt_1 dt_2 dt_3 dt_4 \sum_i u_{a_1 a_2 a_3 a_4}(t_1, t_2, t_3, t_4)
 z_{ia_1}^*(t_1) z_{ia_2}^*(t_2) z_{ia_3}(t_3) z_{ia_4}(t_4) , 
\label{eq:sefftwo}
\end{eqnarray}
where
\begin{eqnarray}
u_{a_1 a_2 a_3 a_4}(t_1,t_2,t_3,t_4)
& = & \frac{1}{4} \int dt_5 dt_6 dt^\prime_5 dt^\prime_6 G^{2c}_{i a_5 a_6 a_6^\prime a_5^\prime}(t_5,t_6,t_6^\prime,t_5^\prime) \nonumber \\
& & \times \left\{ \left[G_{i a_5 a_1}(t_5,t_1)\right]^{-1}
\left[G_{i a_6 a_2}(t_6,t_2)\right]^{-1} \left[G_{i a_3 a_5^\prime}(t_3,t_5^\prime)\right]^{-1}
\left[G_{i a_4 a_6^\prime}(t_4,t_6^\prime)\right]^{-1}  \right. \nonumber \\ & & \left.
 + ( (a_4,t_4) \leftrightarrow (a_3,t_3) ) +
 ( (a_6,t_6) \leftrightarrow (a_5,t_5))  
+ \left((a_6,t_6) \leftrightarrow (a_5, t_5); (a_4,t_4) \leftrightarrow (a_3,t_3)
\right)
\right\}  . \nonumber \\ & &
\end{eqnarray}
Following the 
arguments presented in Appendix B of Ref.~\cite{Dupuis}, it can be shown that
the Green's functions for $z$ are the same as those for the original field $a$.

We note that the following symmetry relations hold for the interaction
kernel $u$ from the definition above:
$$ u_{abcd}(t_1,t_2,t_3,t_4) = u_{bacd}(t_1,t_2,t_3,t_4) = u_{abdc}(t_1,t_2,t_3,t_4).$$
It can also be seen from the definition in Eq.~(\ref{eq:Gcondef}) that
$$G^{2c}_{i a_5 a_6 a_6^\prime a_5^\prime}(t_5,t_6,t_6^\prime,t_5^\prime) 
= G^{2c}_{i a_6 a_5 a_6^\prime a_5^\prime}(t_6,t_5,t_6^\prime,t_5^\prime) 
= G^{2c}_{i a_5 a_6 a_5^\prime a_6^\prime}(t_5,t_6,t_5^\prime,t_6^\prime).$$
\end{widetext}
Similar symmetry relations in the Keldysh structure of four point 
functions were noted in Refs.~\cite{Grass2,Grassthesis}.
Hence there are only 8 independent components we need to evaluate:
$G^{2c}_{qqqq}$, $G^{2c}_{cqqq}$, $G^{2c}_{qqqc}$, $G^{2c}_{qqcc}$, $G^{2c}_{ccqq}$,
$G^{2c}_{cqcq}$, $G^{2c}_{qccc}$ and $G^{2c}_{cccq}$. The remaining four point
function $G^{2c}_{cccc} = 0$ by causality \cite{Chou}.
Explicit expressions for each of the non-trivial components are
written down in Appendix~\ref{app:fourpoint}.  We will find that for our
study of the simplified equations of motion away from the degeneracy points
of the Mott lobes that we will only require $G^{2c}_{cqqq}$, but the 
expressions we provide in Appendix~\ref{app:fourpoint} allow for a more general study of the 
equations of motion than we provide here.

The mean field phase boundary can be determined from the effective
action Eq.~(\ref{eq:sefftwo}) from the vanishing of the coefficient of 
$z_q^* z_{c}$ by noting that
$$ \left<\psi_{i b_1}(t_1) \psi^*_{i b_2}(t_2) \right> = 
-i\tau^1_{b_1 a_1} \tau^1_{b_2 a_2} \left[G_{i a_2 a_1}(t_2,t_1)\right]^{-1},$$
and that the matrix Green's function takes the form
$$\hat{G}(t_1,t_2) = \left(\begin{array}{cc} 0 & {\mathcal G}^A_0(t_1,t_2)
\\ {\mathcal G}^R_0(t_1,t_2) & {\mathcal G}^K_0(t_1,t_2) \end{array}\right),$$
where ${\mathcal G}_0^R$, ${\mathcal G}_0^K$, and ${\mathcal G}_0^A$ are the 
retarded, Keldysh, and advanced Green's functions determined using the
single site Hamiltonian 
$\hat{H}_0$ respectively.  These Green's functions are discussed in more
detail in Appendix \ref{app:aside}.  We can thus obtain
$$\hat{G}^{-1}(t_1,t_2) = \left(\begin{array}{cc} \left[{\mathcal G}_0^{-1}\right]^K(t_1,t_2)
& \left[{\mathcal G}_0^{-1}\right]^R(t_1,t_2)
\\ \left[{\mathcal G}_0^{-1}\right]^A(t_1,t_2) & 0  \end{array}\right),$$
where
\begin{eqnarray}
 \left[{\mathcal G}_0^{-1}\right]^R(t_1,t_2) &= & 
\left[{\mathcal G}_0^R(t_1,t_2)\right]^{-1} , \\
 \left[{\mathcal G}_0^{-1}\right]^A(t_1,t_2) 
& = & \left[{\mathcal G}_0^A(t_1,t_2)\right]^{-1}, \\
 \left[{\mathcal G}_0^{-1}\right]^K(t_1,t_2)&  =  &
- \int dt^\prime dt^{\prime\prime} 
\left[{\mathcal G}_0^R(t_1,t^\prime)\right]^{-1} \nonumber \\
& & \hspace*{0.5cm} \times
{\mathcal G}_0^K(t^\prime,t^{\prime\prime}) 
\left[{\mathcal G}_0^A(t^{\prime\prime},t_2)\right]^{-1} , \nonumber \\ & &
\end{eqnarray}
which along with
${\mathcal G}_0^R(t_1 - t_2) = {\mathcal G}_0^A(t_2 - t_1),$ allows one
obtain the standard equation for the 
mean field phase boundary [Eq.~(\ref{eq:boundary})].

\section{Equations of motion}
\label{sec:eqnmotion}

We can obtain the equations of motion for the order parameter from the saddle 
point conditions on the action:

$$ \frac{\delta S_{\rm eff}}{\delta z_{iq}^*(t)} = 0; 
\quad \quad \frac{\delta S_{\rm eff}}{\delta z_{ic}^*(t)} = 0$$
It is helpful to note that
$$ \left[G_{cc}(t_1,t_2)\right]^{-1} = 0; \quad \left[G_{qq}(t_1,t_2)\right]^{-1} = 
\left[{\mathcal G}_0^{-1}\right]^K(t_2,t_1), $$
$$ \left[G_{qc}(t_1,t_2)\right]^{-1} = \left[{\mathcal G}_0^R(t_1,t_2)\right]^{-1}
 = \left[{\mathcal G}_0^A(t_2,t_1)\right]^{-1},$$
and
$$\left[G_{cq}(t_1,t_2)\right]^{-1} = \left[{\mathcal G}_0^A(t_1,t_2)\right]^{-1}
 = \left[{\mathcal G}_0^R(t_2,t_1)\right]^{-1},$$
to obtain the equations of motion as

\begin{widetext}
\begin{eqnarray}
0 & = & 2J_{ij}(t) z_{j c}(t) + \int_{-\infty}^\infty 
dt_2 \left[{\mathcal G}_0^R(t,t_2)\right]^{-1} z_{i c}(t_2)
+ \int_{-\infty}^\infty dt_2 \left[{\mathcal G}_0^{-1}\right]^K(t,t_2)
 z_{i q}(t_2) \nonumber \\ &  &
+ \frac{1}{2} \int dt_2 dt_3 dt_4 u_{qa_2a_3a_4}(t,t_2,t_3,t_4)
z_{ia_2}^*(t_2) z_{ia_3}(t_3) z_{ia_4}(t_4)  ,
\label{eq:eqnmot1}
\\
0 & = & 2J_{ij}(t) z_{j q}(t) +  \int_{-\infty}^\infty
dt_2 \left[{\mathcal G}_0^A(t,t_2)\right]^{-1} z_{i q}(t_2)
+ \frac{1}{2} \int dt_2 dt_3 dt_4 u_{ca_2a_3a_4}(t,t_2,t_3,t_4)
z_{ia_2}^*(t_2) z_{ia_3}(t_3) z_{ia_4}(t_4) , \nonumber  \\ & &
\label{eq:eqnmot2}
\end{eqnarray} 
with implied summation over $a_2$, $a_3$ and $a_4$. 
 The solution of
these two equations is rather involved in the general case, but the
expressions above allow for the description of the spatial and 
temporal evolution of the superfluid order parameter in both the 
superfluid and Mott insulating phases.  By taking appropriate 
variations of the effective action Eq.~(\ref{eq:sefftwo}) one
may also obtain equations of motion for correlations of the $z$
fields.  In order to gain some insight into the out of equilibrium
dynamics of the situation in which the hopping $J$ is time dependent 
and there is a sweep across the boundary of the superfluid, we
derive a simplified equation for the dynamics of the superfluid
order parameter and study its properties numerically below.

\end{widetext}

\subsection{Simplified equation of motion}
To investigate the nature of the solutions of the equations of
motion, we make some simplifications to 
Eqns.~(\ref{eq:eqnmot1}) and (\ref{eq:eqnmot2}).  We 
focus on low frequencies and long length
scales to determine an equation for the mean field dynamics of the order parameter.

We assume that in the limit $t \to -\infty$, the system is 
in the superfluid phase and the hopping $J(t)$ is
not changing with time.  The initial conditions require
$z_1 = z_2$, which implies that initially $z_q = 0$ and $z_c = \sqrt{2}z_1$,
where $z = z_1$ is the superfluid order parameter.
If $z_q$ remains small under evolution with time
then we can focus only the equation of motion for $z_c$: 
Eq.~(\ref{eq:eqnmot1}).  To see that this is indeed the case, 
we need to note that (see Appendix \ref{app:aside})

\begin{widetext}
\begin{eqnarray}
{\mathcal G}_0^K(\omega) & = & - \frac{2i\pi}{Z} \sum_{r=0}^\infty
e^{-\beta(E_r -\mu r)} \left[(r+1) \, \delta(\omega + \mu -Ur)
 + r\, \delta(\omega + \mu - U(r-1))\right]. \nonumber
\end{eqnarray}
\end{widetext}
Hence terms involving ${\mathcal G}_0^K$ will only contribute to
the low frequency dynamics when $\mu \sim Ur$ for some integer $r$. 
These values of $\mu$ correspond to the values of chemical potential
where for $J=0$ there is degeneracy between Mott insulating states
with $r$ and $r-1$ particles per site.
We restrict ourselves to values of the chemical potential away
from degeneracy, in which case we only need to retain 
terms involving ${\mathcal G}_0^R$ and ${\mathcal G}_0^A$.
In order for $z_q$ to become appreciable, the 
term $u_{cccc}(t,t_2,t_3,t_4)z_c^*(t_2)z_c(t_3) z_c(t_4)$
in Eq.~(\ref{eq:eqnmot2}) must be appreciable.  This term 
depends on the two particle connected Green's function $G^{2c}_{qqqq}$.  
Similarly to ${\mathcal G}_0^K$, $G^{2c}_{qqqq}$
only contributes to low frequency dynamics when $\mu \sim Ur$ for some integer $r$. 
We can thus safely ignore $z_q$ and focus solely on the dynamical equation for
$z_c$: Eq.~(\ref{eq:eqnmot1}).  Taking into account considerations 
about which terms are important for low frequency dynamics as we did
above, it turns out that for values of the chemical potential 
away from $\mu \sim Ur$, the only connected function that we need to evaluate 
is $G^{2c}_{cqqq}$, which is specified in Appendix \ref{app:fourpoint}.  
Writing $z_1 = z$, we can obtain a simplified form of Eq.~(\ref{eq:eqnmot1}) by first noting that

\begin{widetext}
\begin{eqnarray}
\int_{-\infty}^\infty dt_2 \left[{\mathcal G}_0^R(t,t_2)\right]^{-1} z(t_2)
 & = & 
\int_{-\infty}^\infty \frac{d\omega}{2\pi} e^{-i\omega t} \left[{\mathcal G}_0^R\right]^{-1}(\omega)
z(\omega) ,
\end{eqnarray}
which we can expand using 
$$ \left[{\mathcal G}_0^{R}\right]^{-1}(\omega) = \left[{\mathcal G}_0^R\right]^{-1}_{\omega=0} 
+ \omega \left. \frac{\partial}{\partial \omega} \left[{\mathcal G}_0^R\right]^{-1}\right|_{\omega = 0}
 + \frac{1}{2} \omega^2 \left.  \frac{\partial^2}{\partial \omega^2} \left[{\mathcal G}_0^R\right]^{-1} \right|_{\omega = 0}
 + \ldots , $$
leading to
\begin{eqnarray}
\int_{-\infty}^\infty \frac{d\omega}{2\pi} e^{-i\omega t} \left[{\mathcal G}_0^R\right]^{-1}(\omega) z(\omega)
& \simeq & \nu z(t) - i \lambda \frac{\partial z}{\partial t} 
- \kappa^2 \frac{\partial^2 z}{\partial t^2}  ,
\end{eqnarray}
where 
$$ \nu =  \left[{\mathcal G}_0^R\right]^{-1}_{\omega=0}; \quad
\lambda = - \left. \frac{\partial}{\partial \omega} \left[{\mathcal G}_0^R\right]^{-1} \right|_{\omega = 0} ; \quad
\kappa^2 = \frac{1}{2}  \left. \frac{\partial^2}{\partial \omega^2} 
\left[{\mathcal G}_0^R\right]^{-1} \right|_{\omega = 0}  .$$

Explicit expressions for $\nu$, $\lambda$ and $\kappa^2$ can be easily computed from 
Eqs.~(\ref{eq:retfiniteT}) and (\ref{eq:retzeroT}) and are given in Appendix \ref{app:eqmot}.
The temperature and chemical potential dependence of these quantities is displayed 
in Figs.~\ref{fig:parameters} a) - c).  The phase boundary of the superfluid phase
at finite temperature is shown for reference in Fig.~\ref{fig:parameters} d).  We 
can see that the strongest temperature dependence of the parameters is for values of
$\mu/U$ close to an integer (which we ignore), and that both $\lambda$ and $\kappa^2$
are relatively insensitive to thermal effects over a wide range of $\mu/U$ values.
The interaction term $u$ is most sensitive to temperature and starts to deviate 
strongly from its zero temperature value by temperatures as large as $T \simeq 0.2 U$,
which corresponds to the temperature at which there is full melting of the 
insulating phase \cite{Mahmud,Gerbier}.

\begin{figure}[h]
\includegraphics[width=6cm,angle=270]{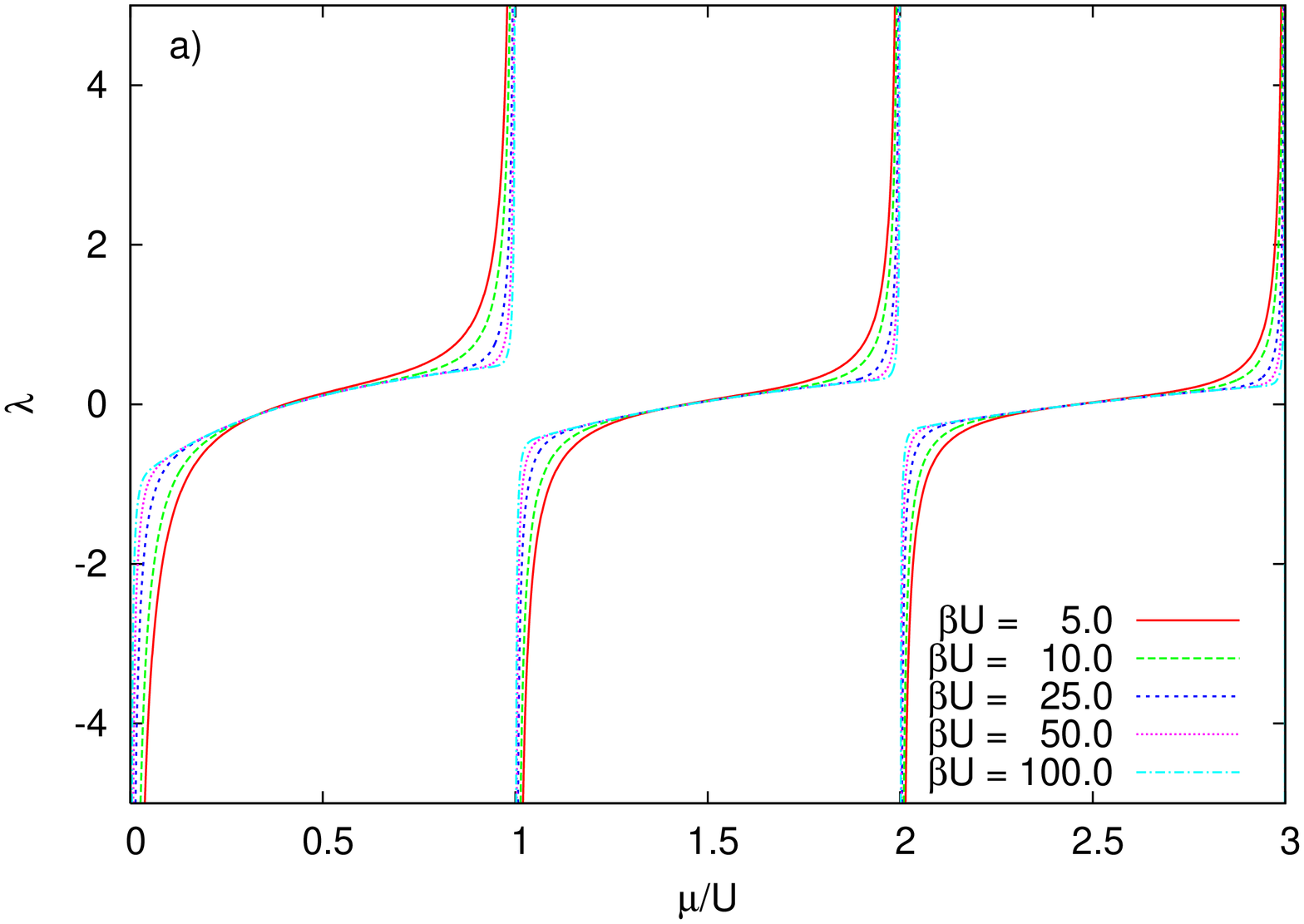}
\includegraphics[width=6cm,angle=270]{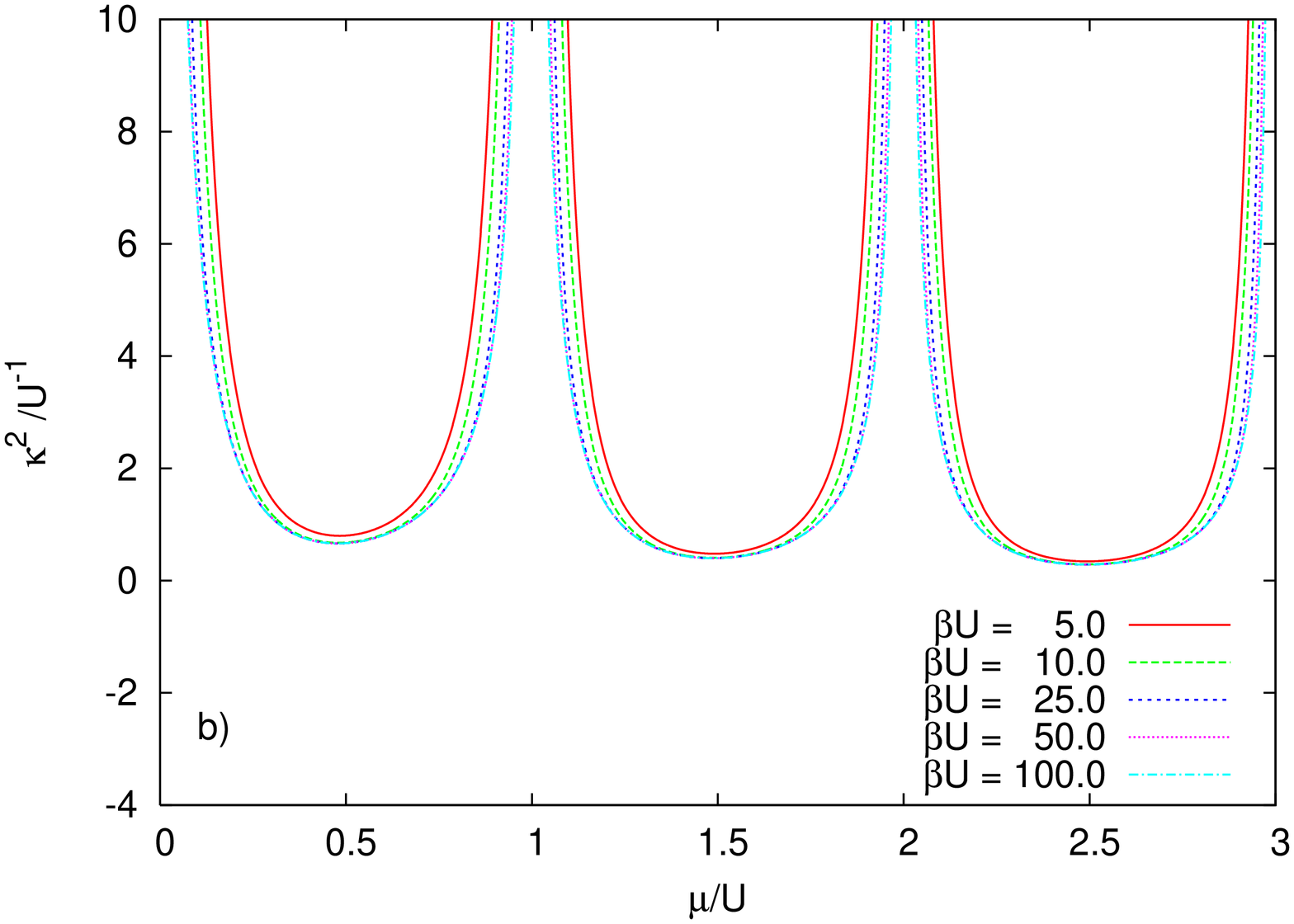}
\includegraphics[width=6cm,angle=270]{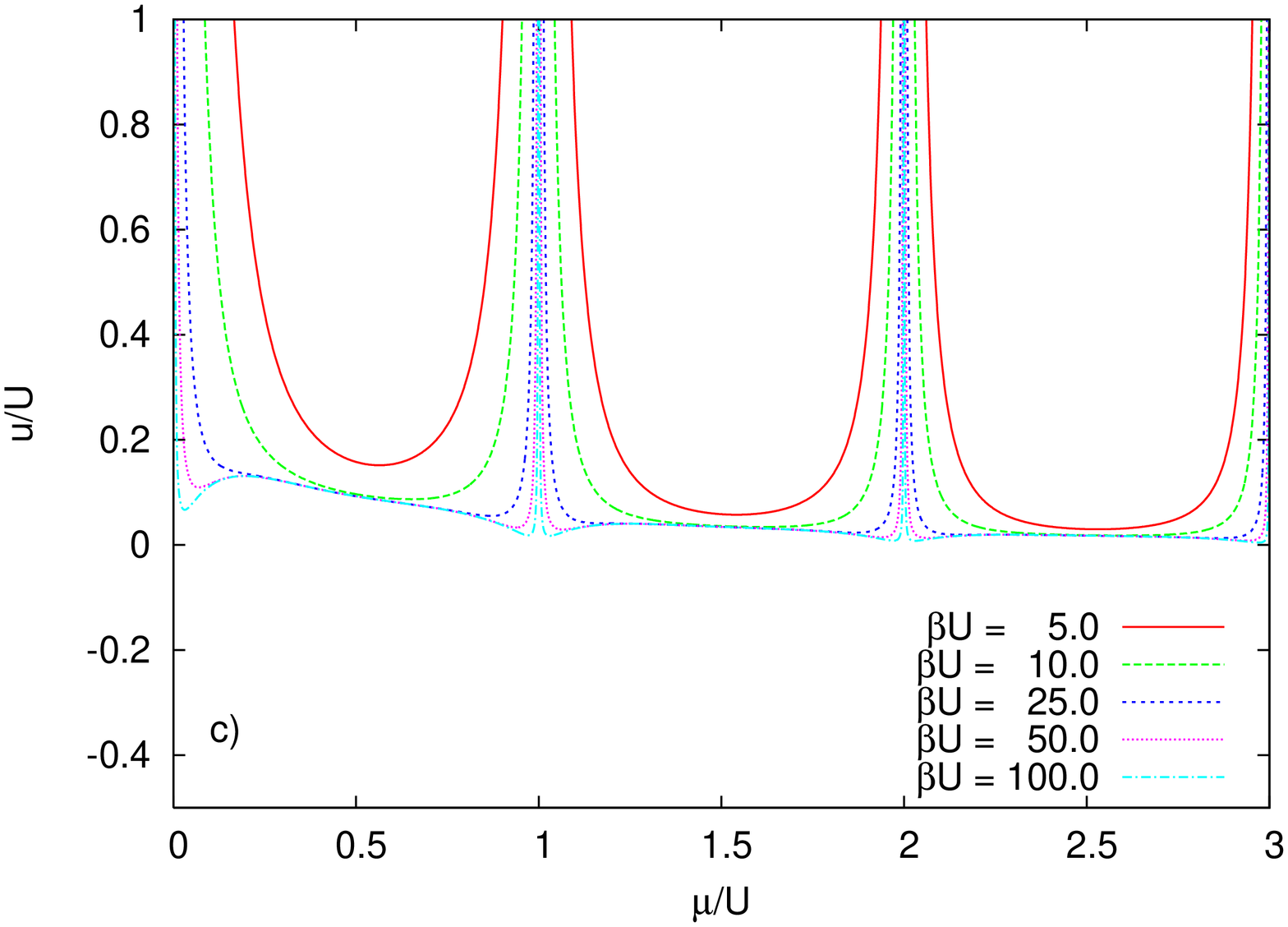}
\includegraphics[width=6cm,angle=270]{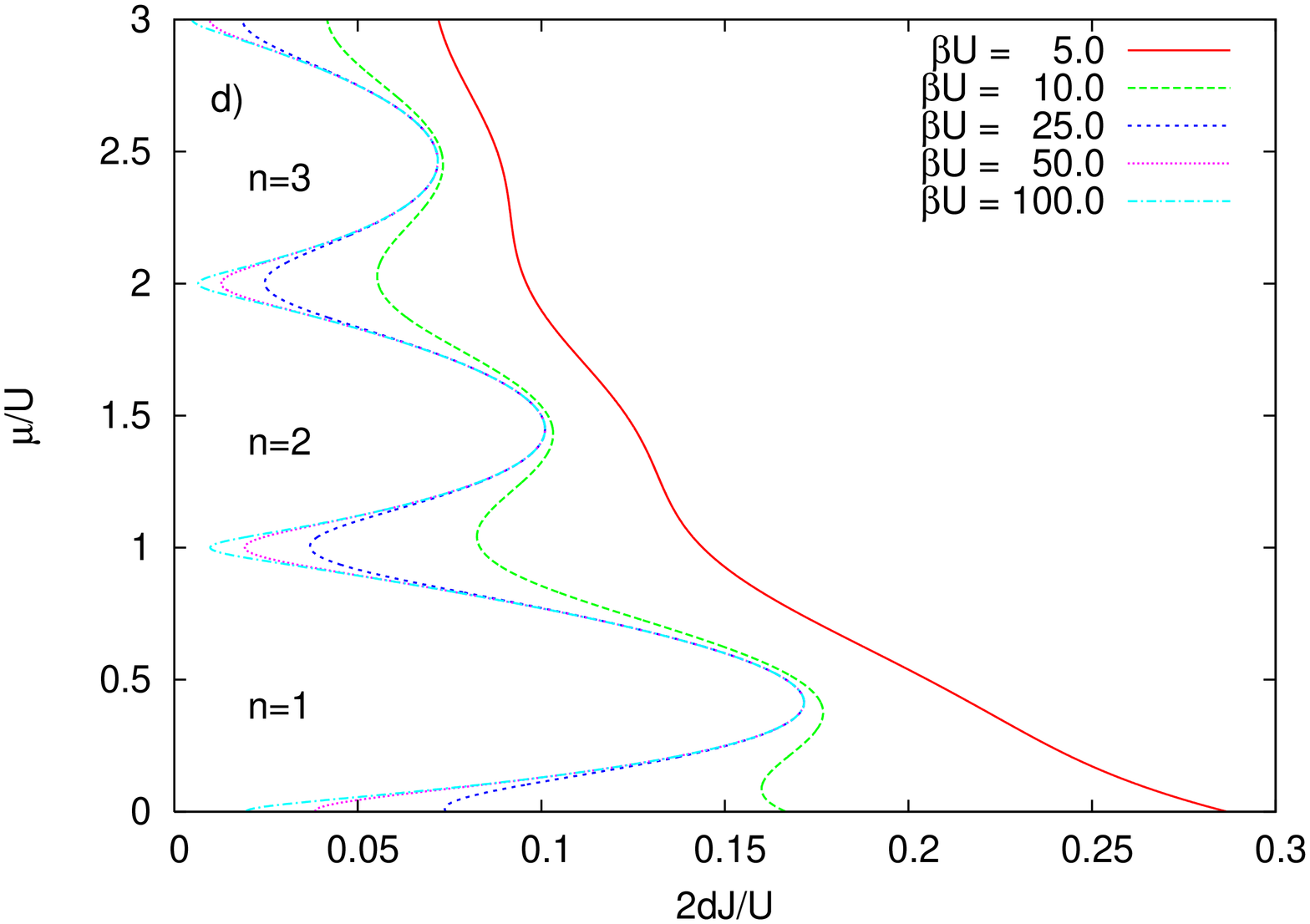}
\caption{Plots of a) $\lambda$; b) $\kappa^2$; and c) $u$ as 
a function of $\mu/U$ and inverse temperature $\beta$. d) The 
phase boundary of the superfluid state is shown in as a 
function of $\mu/U$ and $2dJ/U$ at several temperatures
for reference. The filling per site in the Mott insulating
phase at zero temperature is indicated.}
\label{fig:parameters}
\end{figure}
\end{widetext}
Using a similar expansion to the one used to derive the mean field phase
boundary, we can note that after a Fourier transform in space
\begin{eqnarray}
2J_{ij}(t) z_{jc}(t) & \to & 2J(t) \sum_{j=1}^d \cos(k_j a)\, z(\bvec{k},t) \nonumber
\\ & \simeq & 2J(t)\left[d - \frac{1}{2} k^2 a^2\right] z(\bvec{k},t) ,
\end{eqnarray}
for small $ka$.  We will focus on the long wavelength limit and ignore 
terms of order $ka$.

We only retain the $k=0$ part of the interaction term, in keeping with our
focus on long wavelength physics, and we take the  
low frequency limit of the interaction term by expanding the two particle 
connected Green's function and the retarded and advanced Green's functions 
about the $\omega = 0$ limit.  Recalling from above that $|z_c(t)|^2 = 2|z(t)|^2$, we may 
approximate the interaction term by $-u|z|^2 z$, where $u$ is stated in
Appendix \ref{app:eqmot} and is in accord with the static value
calculated for equilibrium in Ref.~\cite{SenguptaDupuis}.  Thus 
we have as our approximation to the equation of motion:

$$\left[2dJ(t) + \nu\right] z(t) 
 - i\lambda \frac{\partial z(t)}{\partial t} 
- \kappa^2 \frac{\partial^2 z(t)}{\partial t^2} - u|z(t)|^2 z(t) = 0.$$

Take $J(t) = J_0 + j(t)$, where $J_0$ is chosen so that 
$$ 2dJ_0 + \nu = 0,$$
i.e. $J_0$ is chosen to
lie on the mean field phase boundary for the superfluid 
for a given $\mu$.  Hence we 
may write the approximate mean field equation of motion as

\begin{eqnarray}
\kappa^2 \frac{\partial^2 z}{\partial t^2} + 
i\lambda \frac{\partial z}{\partial t} + \delta(t) z + u|z|^2 z = 0,
\label{eq:mfeqnmot}
\end{eqnarray}
where $\delta(t) = -2d j(t)$.  Even after the simplifications
made above, this equation for the dynamics of the order parameter
is a non-linear second order differential equation, for which 
we are not able to find analytic solutions in general.  Below 
we discuss numerical solutions of this equation, along with an
analytic solution that can be determined in a special
case which illuminates the properties of the solutions of the 
equation.  

We study Eq.~(\ref{eq:mfeqnmot}) for fixed $\mu$ and time-varying $J$.
In experiment, there is a confining potential so that there is 
a position dependent local chemical potential $$\mu_{\rm local}(r)
= \mu - V(r),$$
where $V(r)$ is the trapping potential.  The solutions
we obtain for the dynamics at fixed $\mu$ should be compared to the 
experimental situation in which one views the dynamics at fixed
radius in a symmetric trap. (This picture should be reasonable 
at time scales shorter than the timescale for global mass 
redistribution in the trap, which can be quite long compared to 
microscopic timescales \cite{Natu}).

\begin{figure}
\includegraphics[width=8cm]{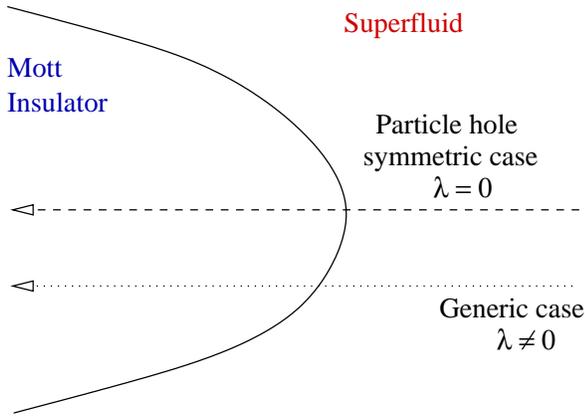}
\caption{Two possibilities for quantum phase transition at constant $\mu$.}
\label{fig:lobe}
\end{figure}

If we fix $\mu$, then there are two possibilities for the dynamics 
that we should consider: a) the particle-hole
symmetric case, in which case $\lambda = 0$, and b) the generic case,
in which $\lambda \neq 0$.  The particle-hole symmetric case 
corresponds to the transition at the tip of the Mott lobe 
as illustrated in Fig.~\ref{fig:lobe}.

We consider traversal of the quantum critical region as $\delta(t)$ 
varies with $t$.  We demand that 
$$\lim_{t \to -\infty} \delta(t) = -\delta_0; \quad {\rm and} \quad 
\lim_{t \to \infty} \delta(t) = \delta_1.$$   
In our numerical solutions we use the form
\beq
\delta(t) = \left(\frac{\delta_0 + \delta_1}{2}\right)
\tanh\left(\frac{t}{\tau_Q}\right) + \frac{\delta_1 - \delta_0}{2} ,
\label{eq:delta}
\eeq
where, similarly to Cucchietti {\it et al.} \cite{Cucchietti}, 
who studied the transition from Mott insulator to superfluid in the
one dimensional BHM, 
we assume that there is a  timescale $\tau_Q$ which is the characteristic time
for $\delta(t)$ to cross from $-\delta_0$ to $\delta_1$.  

\subsection{Particle-hole symmetric case}
In the particle-hole symmetric case, $\lambda = 0$ and 
the saddle point equation takes the form

\beq
\kappa^2 \frac{\partial^2 z}{\partial t^2} + \delta(t)z + u|z|^2 z = 0 .
\label{eq:particlehole}
\eeq
We can choose $z(t) = \rho(t) e^{i\theta(t)}$, and then
real and imaginary parts of the equation give
\beq
0 & = & \ddot{\rho} - \rho\left(\dot{\theta}\right)^2 + \bar{\delta}\rho + u\rho^3, \nonumber \\
0 &= & 2\dot{\theta}\dot{\rho} + \ddot{\theta}\rho \nonumber
\eeq
where we rescaled
$$ t = \kappa\bar{t}; \quad \quad \bar{\delta}(\bar{t}) = \delta(\kappa \bar{t}) = \delta(t),$$
and wrote the equations in terms of the rescaled time co-ordinate $\bar{t}$.
The second equation can be integrated to give
$$ \ln\left(\dot{\theta}\right) = - 2\ln\rho + c_1,$$ i.e. $\dot{\theta}\rho^2 = c,$
so
$$\ddot{\rho} - \frac{c^2}{\rho^3} + \bar{\delta}\rho + u\rho^3 = 0.$$
The initial condition that the system is deep in the superfluid phase 
implies that as $t \to -\infty$, $\dot{\theta} \to 0,$ and $\dot{\rho} \to 0$, so
$\rho = \sqrt{\frac{\delta_0}{u}},$ and $c=0$ (we choose $\theta = 0$
without loss of generality). Thus
$$\ddot{\rho} + \bar{\delta}\rho + u\rho^3 = 0.$$
Rescaling $\rho \to \tilde{\rho}/\sqrt{u}$, then dropping the tilde and bar,
$$\ddot{\rho} + \delta(t)\rho + \rho^3 = 0,$$ 
with $\rho \to \sqrt{\delta_0}$ as $t \to -\infty$.
In the long time limit, when $\delta(t) = \delta_1$,
then we may rewrite the differential equation for $\rho$ as 
$$\frac{d}{dt}\left[\frac{1}{2}\left(\dot{\rho}\right)^2 + \frac{\delta_1}{2}\rho^2 + \frac{1}{4}\rho^4\right] =0 ,$$
Then
$$\dot{\rho}^2 + \delta_1 \rho^2 + \frac{1}{2}\rho^4 = A,$$
and, writing $\rho = \xi y$, $t = \eta x$, we have
$$ \left(\frac{dy}{dx}\right)^2 = (1-k^2) - (1-2k^2) y^2 - k^2 y^4,$$
with $$1-k^2 = \frac{\eta^2A}{\xi^3}, \quad k^2 = \frac{1}{2}\xi^2 \eta^2, \quad 1-2k^2 = \delta_1 \eta^2,$$
and we can solve to get $$ k = \frac{1}{\sqrt{2}} \frac{1}{\sqrt{1 + \frac{\delta_1}{\xi^2}}},$$
which must satisfy $0<k<1$.  The solution to our equation as $t \to \infty$ is thus
$$\rho = \xi {\rm cn}\left(\frac{\xi t}{\sqrt{2} k};k\right),$$
which in the original variables is
$$z(t) = \frac{\xi}{\sqrt{u}} {\rm cn}\left(\frac{\xi}{\sqrt{2}k}\frac{t}{\kappa}; k\right).$$
In general we cannot determine the value of $\xi$ analytically.   We can obtain an analytical
solution if there is  a 
jump in $\delta(t)$ from $-\delta_0$ to $+\delta_1$ at $t=0$. [Note that this form of $\delta(t)$
violates the assumption that we made in deriving the equation that frequencies are low, but the
solution in this case is still instructive, as it shares many 
features with the solution for more physical forms of $\delta(t)$.]
We know $z(t) = \sqrt{\frac{\delta_0}{u}}$ for $t < 0$, and recalling ${\rm cn}(0;k) = 1$, we 
get $\xi = \sqrt{\delta_0}$, which implies 
$$k = \frac{1}{\sqrt{2}} \frac{1}{\sqrt{1 + \frac{\delta_1}{\delta_0}}}, $$ and so we get

\beq
z(t) = \sqrt{\frac{\delta_0}{u}}{\rm cn}\left(
\frac{\sqrt{(\delta_0 + \delta_1)}t}{\kappa};\frac{1}{\sqrt{2}} 
\frac{1}{\sqrt{1 + \frac{\delta_1}{\delta_0}}}\right), \nonumber
\eeq
which is periodic in time with average value 0 and period 
$$4K\left(\frac{1}{\sqrt{2}} \frac{1}{\sqrt{1 + 
\frac{\delta_1}{\delta_0}}}\right) \frac{\kappa}{\sqrt{\delta_0 + \delta_1}}$$ 
where
$$K(k) = \int_0^{\frac{\pi}{2}} \frac{d\theta}{\sqrt{1-k^2\sin^2\theta}}.$$

\begin{figure}
\includegraphics[width=8cm]{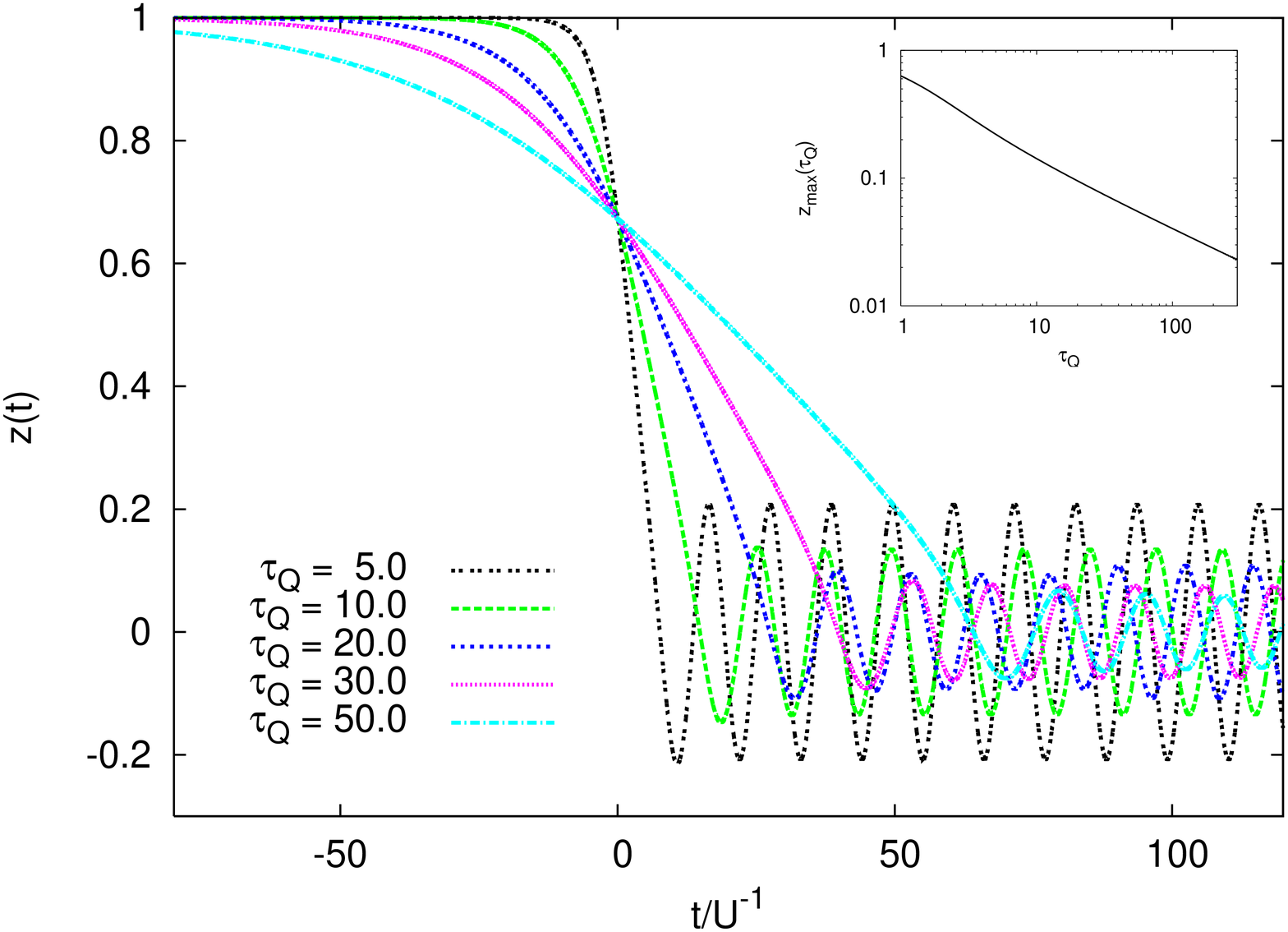}
\caption{Dynamics of $z(t)$ normalized to unity in the 
particle hole symmetric case, for a variety of $\tau_Q$.
The parameters are $\beta U = 100$,
$\mu = 0.4142136$, $\kappa^2 = 0.707107$, $u = 0.1038$,
and we take $\delta_0 = 1.83$, $\delta_1 = 0.17 = J_0(\mu)$.  This
corresponds to a quench from $2dJ/U = 2.0$ to $2dJ/U = 0.0$. 
The inset shows the value of $z_{\rm max}(\tau_Q)$ as a function of $\tau_Q$.}
\label{fig:parthole}
\end{figure}

We obtain numerical solutions of Eq.~(\ref{eq:particlehole}) with $\delta(t)$ taking the
form given in Eq.~(\ref{eq:delta}) at the particle-hole symmetric 
point in the first Mott lobe for several different values of $\tau_Q$,
as displayed in Fig.~\ref{fig:parthole}. One can see that in each case, for large
values of $t \gg \tau_Q$, the form of the solution is that $z(t)$ oscillates 
in a periodic manner with a magnitude that decreases with increasing $\tau_Q$.
When averaged over a period $T$ at times $t \gg \tau_Q$,
$$\left<z\right>_T = \frac{1}{T}\int_t^{t+T} d\tilde{t} \, z(\tilde{t}) = 0,$$
as we would expect in the Mott insulating state.
Defining $z_{\rm max}(\tau_Q) = \lim_{t \to \infty} |z(t)|$ we can see that
$z_{\rm max}(\tau_Q)$ decreases with increasing $\tau_Q$ without any
indication of saturation, as illustrated in the inset to Fig.~\ref{fig:parthole}.
Note that in our numerical simulations $t$ is measured
in units of $U^{-1}$.

\subsection{Generic case}
\begin{figure}
\includegraphics[width=6cm,angle=270]{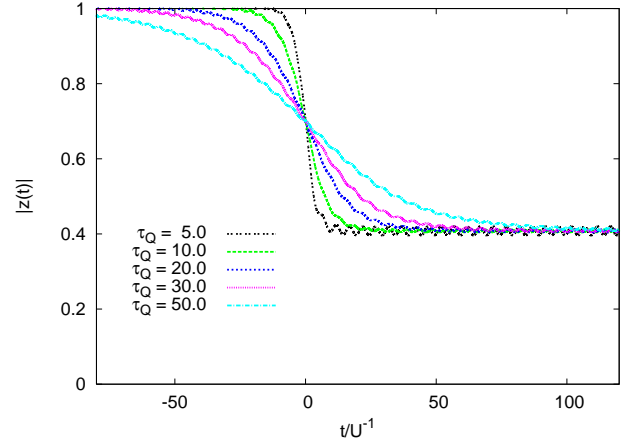}
\caption{Dynamics of $|z(t)|$ normalized to unity in the
generic case, for a variety of $\tau_Q$.
The parameters are $\beta U = 100$,
$\mu = 0.25$, $\lambda = -0.28$, $\kappa^2 = 1.55$, $u = 0.1277$,
and we take $\delta_0 = 1.85$, $\delta_1 = 0.15 = J_0(\mu)$.  This
corresponds to a quench from $2dJ/U = 2.0$ to $2dJ/U = 0.0$.}
\label{fig:gentrace}
\end{figure}

In the generic case in which $\lambda \neq 0$, we 
start with Eq.~(\ref{eq:mfeqnmot}) and try 
for a solution of the form $$z(t) = \rho(t) e^{i\theta(t)}.$$
Taking real and imaginary parts of the equation, after substitution 
gives
\begin{eqnarray}
\kappa^2\left(\ddot{\rho} - \rho \left(\dot{\theta}\right)^2\right) -\lambda \rho \dot{\theta}
 + \delta(t)\rho + u\rho^3  & = & 0, \nonumber \\
\kappa^2\left(2\dot{\theta}\dot{\rho} + \ddot{\theta}\rho\right) + \lambda \dot{\rho} & = & 0. 
\nonumber
\end{eqnarray}
Integrating the second equation with respect to $t$ leads to
\begin{eqnarray}
\dot{\theta} = \frac{c - \frac{\lambda}{2}\rho^2}{\kappa^2 \rho^2},
\end{eqnarray}
In the $t\to -\infty$ limit, $\rho$ and $\theta$ are constant, so 
we can determine
$c = \frac{\lambda}{2}\rho_0^2 = \frac{\lambda \delta_0}{2u},$
and we obtain the following equation for $\rho$:
\begin{eqnarray}
\kappa^2 \ddot{\rho} - \frac{\lambda^2}{4\kappa^2 \rho^3} 
\left(\frac{\delta_0}{u}\right) + \delta(t) \rho + u \rho^3 = 0 .
\label{eq:gensim}
\end{eqnarray}
We solve Eq.~(\ref{eq:gensim}) numerically for a variety of 
values of $\tau_Q$ and display $|z(t)| = \rho(t)$ for $\mu/U = 0.25$
(well away from both degeneracy and the particle hole symmetric case)
in Fig.~\ref{fig:gentrace}. 

\begin{figure}
\includegraphics[width=6cm,angle=270]{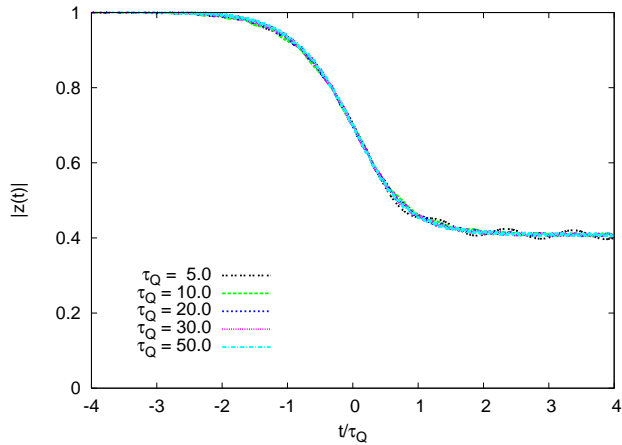}
\caption{Dynamics of $|z(t)|$ normalized to unity in the
generic case, for a variety of $\tau_Q$ with time rescaled
by $\tau_Q$. The parameters are as in Fig.~\ref{fig:gentrace}.}
\label{fig:collapse}
\end{figure}

The solution displays the similar
feature to the particle-hole symmetric case that the average of $z$
over a period $\left<z\right>_T =0$.  However, it is clear that 
as $\tau_Q$ increases, there does not seem to be any decay in the 
values of $|z(t)|$.  By rescaling the time with $\tau_Q$, we can see
that in fact the different traces collapse onto each other, as
we display in Fig.~\ref{fig:collapse}.

\subsection{Chemical potential and temperature dependence of dynamics}
\begin{figure}
\includegraphics[width=6cm,angle=270]{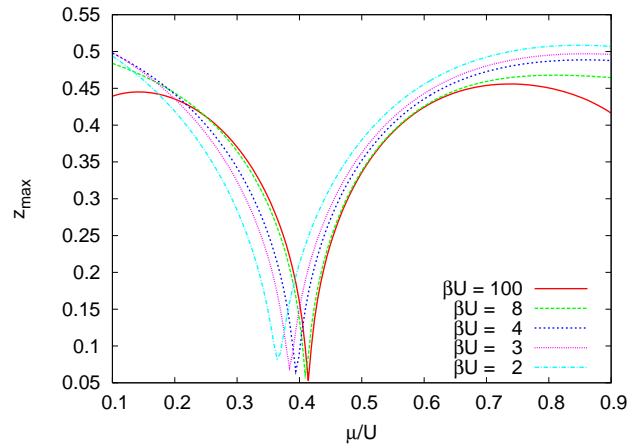}
\caption{Limit of $z_{\rm max}$ in the large $\tau_Q$ limit at several
temperatures.}
\label{fig:phimaxtau}
\end{figure}

The traces of $z(t)$ and $|z(t)|$  that we displayed in 
Figs.~\ref{fig:parthole}-\ref{fig:collapse} were for a particular value
of the chemical potential in the generic case and for a low temperature 
($\beta U = 100$) in both cases.  It is of interest to see whether 
the observation that in the non particle-hole symmetric case that 
there is a metastable state after a quantum quench is robust to 
variations of chemical potential and temperature.  Defining 
$z_{\rm max} = \lim_{\tau_Q \to \infty} z_{\rm max}(\tau_Q)$, 
we calculated this for
$0.1 < \mu/U < 0.9$ and temperatures ranging from $\beta U = 100$
to $\beta U = 2$.  We focus only on the first Mott
lobe, but from perusal of the chemical potential and temperature
dependence of the parameters $\lambda$, $\kappa^2$ and $u$ in 
Fig.~\ref{fig:parameters}, we expect that similar qualitative 
results should be obtained for other Mott lobes.  We find that
apart from the particle-hole symmetric point, where we believe
the displayed finite value of $z_{\rm max}$ is an artefact of
our numerical calculations, that the transition to a metastable
state in which $z_{\rm max} \neq 0$ is generic for a wide 
range of values of $\mu$ and persists to temperatures 
comparable to the melting temperature of the insulator 
as illustrated in Fig.~\ref{fig:phimaxtau}.  It should be 
noted that the physics that we have left out of dynamical
equation, namely spatial dependence of $z$ and also 
higher frequency components of $z$ will presumably lead
to equilibration of $z$ at long enough times, but
as we argue in Sec.~\ref{sec:disc},  it 
may well be reasonable to expect that the behaviour 
we identify at the mean field level to be experimentally relevant
on appropriate timescales.

\section{Discussion and Conclusions}
\label{sec:disc}
In this paper we have derived a real time effective action for the
Bose Hubbard model using the Schwinger-Keldysh technique,
generalizing previous work that obtained an equilibrium effective
action \cite{SenguptaDupuis}.  This
action allows for a description of the properties of both the 
superfluid and Mott insulating phases.  Hence we are able to
study the out of equilibrium dynamics as the parameters in the
Hamiltonian are changed so that the ground state is tuned from
one phase to another.  We obtain the saddle point equations of motion and
by focusing on low frequency, long wavelength dynamics are 
able to obtain an equation of motion for the superfluid order
parameter.  We have focused on this case as the simplest 
example of dynamics, but we emphasise that our approach leads 
to equations of motion that can be used to study high 
frequencies and spatial variations of the order parameter and
its correlations.

We study the equations of motion by varying the hopping parameter
$J$ as a function of time at fixed chemical potential 
to sweep from deep in the superfluid 
phase to deep in the Mott insulating phase over a timescale of 
order $\tau_Q$.  We study the $\tau_Q$ dependence of the superfluid
order parameter numerically and find that in the long $\tau_Q$ limit
the system generically reaches a state in which the time averaged
value of the order parameter is zero (as would be expected in 
equilibrium for a Mott insulator), but the absolute value of the 
order parameter is non-zero.  The magnitude of the order parameter
in the long $\tau_Q$ limit appears to vanish only at the particle-hole
symmetric value of the chemical potential, and grows with distance
from the particle-hole symmetric value of $\mu$.  The generic final 
state is clearly an out-of-equilibrium metastable state, with 
equilibration only possibly for the particle-hole symmetric case.
The generically non-zero value of $|z(t)|$ in the final state 
indicates that the system retains memory of the initial superfluid
state, a feature which is observed in quantum revival experiments 
\cite{Greiner2,Sebby,Will}
that indicate quantum coherence remains even after a quench into 
the insulating phase.

There have been several other recent theoretical works on the 
out-of-equilibrium dynamics of the Bose Hubbard model that see
evidence of the system entering a metastable state after a 
sweep from the superfluid phase to the Mott insulating state.
Sch\"{u}tzhold {\it et al.} \cite{Schutzhold} studied the dynamics
in the limit of large number of bosons per site and found a slow
decay of the superfluid fraction for a slow sweep from the 
superfluid phase to the Mott insulating phase.  Kollath {\it et al}.
\cite{Kollath} investigated the one and two dimensional BHM numerically with 
the number of bosons fixed to an average of one boson per site
and found that for small enough values of the final value of the 
hopping, the system reached a non-thermal steady state which 
was relatively insensitive to the details of the initial state.
These authors determined whether the system was thermal or not
by investigating real-space correlations, so it is not possible
to make a direct comparison with our results here.  Most recently
Sciolla and Biroli \cite{Sciolla} considered the infinite 
dimensional Bose Hubbard model at integer filling and also found 
that the final state after a quantum quench of $U$ showed a non-zero 
superfluid order parameter.  Similar features have also been 
reported for mean field studies of fermions after a quantum quench \cite{Schiro}.

Whilst the emergence of a metastable state after a quench from 
the superfluid to the insulating state is also seen in our 
work, we study a different situation to the previous 
works.  We consider a spatially uniform BHM, as do 
Refs.~\cite{Schutzhold,Kollath,Sciolla}, but we consider fixed
chemical potential rather than fixed particle number.  To
compare  theoretical descriptions of the out of 
equilibrium dynamics of the Bose Hubbard model and experiments
on the quench dynamics of a fixed number of cold atoms in an optical lattice, 
the physical meaning of working with fixed chemical potential 
needs to be discussed.  The presence of a spatially non-uniform
trapping potential means that instead of viewing the system as 
having a uniform chemical potential, it is often more convenient
to view the system as having a spatially dependent {\it local
chemical potential}:  $\mu_{\rm local}(\bvec{r}) = \mu - V(\bvec{r})$,
where $V(\bvec{r})$ is the trapping potential.  For a symmetric 
trap, this implies that a reasonable description of the phase the system
is in at radius $r$ can be determined by using $\mu_{\rm local}(r)$ 
-- this implies the ``wedding cake'' structure seen in many 
experiments.  Our study of the equations of motion at fixed
chemical potential would then correspond to studying the dynamics
of atoms in a trap at fixed radius (albeit with radii corresponding
to certain values of the chemical potential excluded due to the 
approximations we made in deriving the equation of motion).

This viewpoint appears to be borne out in recent experiments 
\cite{Bakr,Hung,Sherson,Chen} and theoretical work 
\cite{Natu,Bernier} on quantum quenches for cold bosons.
Natu {\it et al}. \cite{Natu} argue that the very large differences in 
relaxation times observed in Refs.~\cite{Bakr} (of order ms) 
and \cite{Hung} (of order $\sim$ 1s)
can be understood if one looks at mass transport during 
equilibration.  If the average number of particles per site
remains the same in crossing from the superfluid to an insulator,
then equilibration can be quick as in Ref.~\cite{Bakr}, but if the average number of 
particles per site needs to change, then there must be mass
transport and the equilibration is slow as in Ref.~\cite{Hung}.  The results we find 
for the long time limit of $z_{\rm max}$ illustrated in 
Fig.~\ref{fig:phimaxtau} are in accord with this idea.  For
the chemical potential associated with particle hole
symmetry, the value of $z_{\rm max}$ decays to (close to) zero,
whereas for other values of $\mu$, $z_{\rm max}$ can be an 
appreciable fraction of the value of $|z|$ in the initial state.
At the particle hole symmetric $\mu$, the average number of 
bosons per site does not change in crossing from the superfluid
to the Mott insulator \cite{Fisher}, in accord with the condition
for local equilibration without mass transport \cite{Natu}.  
Global mass transport is not captured within our simplified
equation of motion, and there is no decay of the metastable state
and equilibration on a longer time scales. 

The main results of our work and their connection to existing
experimental and theoretical work in the field of cold atoms
are outlined above, but there are a number of future directions
that it might be interesting to pursue based on what we have 
done here.  First, a more thorough study of the solutions of
the equations of motion allowing for spatial fluctuations and
higher frequencies than we consider here might lead to further
insight into the dynamics of the Bose Hubbard model.  The 
inclusion of a trapping potential would also allow for additional
contact with experiment \cite{Buchhold}.  Second, it would be interesting to 
add the effects of dissipation \cite{Dalidovich,Robertson}, which
has been shown to renormalize the phase boundaries in the BHM.
For cold atoms the effects of dissipation can probably be 
ignored, but in other realizations of the BHM this may not be feasible
\cite{Tomadin}.

Recent experimental advances which allow for high spatial resolution
in cold atom experiments \cite{Gemelke,DeMarco,Sherson,Ma,Bakr,Weitenberg} suggests that there
will be advanced capabilities for probing the out of equilibrium
dynamics spatially as well as temporally, suggesting that there are
exciting times ahead for studies of out of equilibrium dynamics 
of Bose Hubbard systems.

\section{Acknowledgements}
The authors thank Claudio Chamon for helpful discussions 
and encouragement, especially in the early stages of this work
and Jeff McGuirk for a critical reading of the manuscript.
This work was supported by NSERC.

\begin{appendix}
\begin{widetext}

\section{Hubbard-Stratonovich transformation}
\label{app:HS}

Starting from the identities (where $z = x+iy$)
$$ \int_{-\infty}^\infty \int_{-\infty}^\infty \frac{dx dy}{i\pi} e^{i|z|^2}
 = \int_{-\infty}^\infty \int_{-\infty}^\infty \frac{dx dy}{(-i\pi)} 
e^{-i|z|^2} = 1,$$
it is easy to show that
$$ e^{-ia^* a} = \int_{-\infty}^\infty \int_{-\infty}^\infty \frac{dx dy}{i\pi}
e^{i|z|^2 + i(z^* a + z a^*)}; \quad 
e^{ia^* a} = \int_{-\infty}^\infty \int_{-\infty}^\infty 
\frac{dx dy}{(-i\pi)} e^{-i|z|^2 + i(z^* a + z a^*)}.$$
Using these results we may write (with $z_1 = x_1 + iy_1$,  $z_2 = x_2 + iy_2$
and $z_3 = x_3 + iy_3$)
\begin{eqnarray}
e^{-i(\xi^*\eta + \eta^* \xi)}
& = & e^{-i(\xi^* + \eta^*)(\xi + \eta) + i\xi^*\xi + i\eta^* \eta} \nonumber 
\\
& = & \int \frac{dx_1 dy_1}{i\pi} \int \frac{dx_2 dy_2}{(-i\pi)}
\int \frac{dx_3 dy_3}{(-i\pi)} e^{i|z_1|^2 - i|z_2|^2 - i|z_3|^2}
e^{i\left(z_1(\xi^* + \eta^*) + z_1^*(\xi + \eta) + z_2 \xi^* + z_2^* \xi + z_3 \eta^* + z_3^* \eta\right)} \nonumber \\
& = & \int \frac{d\tilde{x}_1 d\tilde{y}_1}{i\pi} \frac{d\tilde{x}_2 d\tilde{y}_2}{(-i\pi)} 
\frac{d\tilde{x}_3 d\tilde{y}_3}{(-i\pi)} e^{i|\tilde{z}_1|^2 -i|\tilde{z}_2 - \tilde{z}_1|^2 - 
i |\tilde{z}_3 - \tilde{z}_1|^2} e^{i\left(\tilde{z}_2^* \xi + \tilde{z}_2 \xi^* + \tilde{z}_3^* \eta
 + \tilde{z}_3 \eta^*\right)} \nonumber
\end{eqnarray}
where we change variables to $\tilde{z}_1 = z_1$, $\tilde{z_2} = z_1 + z_2$, 
and $\tilde{z}_3 = z_1 + z_3$.  After integrating out $\tilde{z}_1$, we get
\begin{eqnarray}
e^{-i(\xi^* \eta + \xi \eta^*)} & = & \int \frac{d\tilde{x}_2 d\tilde{y}_2}{(-i\pi)} 
\frac{d\tilde{x}_3 d\tilde{y}_3}{i\pi} e^{2i(\tilde{x}_2 \tilde{x}_3 + \tilde{y_2} \tilde{y}_3)}
e^{i(\tilde{z}_2^* \xi + \tilde{z}_2 \xi^* + \tilde{z}_3 \eta^* + \tilde{z}_3^* \eta)} \nonumber \\
& = & \int \overline{\mathcal D}(z_2,z_2^*){\mathcal D}(z_3,z_3^*) e^{i(z_2^* z_3 + z_2 z_3^*)}
e^{i(z_2^* \xi + z_2 \xi^* + z_3 \eta^* + z_3 \eta)}, 
\end{eqnarray}
where $${\mathcal D}(z,z^*) = \frac{dx dy}{i\pi}; \quad \quad \overline{\mathcal D}(z,z^*) = \frac{dx dy}{(-i\pi)}.$$
\end{widetext}

\section{Mean field phase boundary}
\label{app:aside}
One way to determine the mean field 
phase boundary between the superfluid and Mott insulating 
phases is to determine when the coefficient of the quadratic term in the action 
Eq.~(\ref{eq:firstHS})
vanishes.  In order to do this it is helpful to note that

\begin{eqnarray}
 \tau^1_{a_1 b_1} G_{ib_1 b_2}(t_1,t_2)
\tau^1_{b_2 a_2} = \left(\begin{array}{cc} {\mathcal G}_0^K(t_1,t_2) & 
{\mathcal G}_0^R(t_1,t_2) \\ {\mathcal G}_0^A(t_1,t_2) & 0 \end{array} \right),
\nonumber \\
\end{eqnarray}
where ${\mathcal G}_0^R$, ${\mathcal G}_0^A$, and ${\mathcal G}_0^K$
are the retarded, advanced and Keldysh propagators respectively, with the 
subscript $0$ indicating that these are the propagators associated with $H_0$.
The definitions of the propagators are:

\beq
i{\mathcal G}_0^K(t-t^\prime) & = & i{\mathcal G}_0^<(t,t^\prime)
+ i{\mathcal G}_0^>(t,t^\prime) \nonumber , \\
i{\mathcal G}_0^R(t-t^\prime) & = & \theta(t-t^\prime)
[i{\mathcal G}_0^>(t,t^\prime) - i{\mathcal G}_0^<(t,t^\prime)], \nonumber \\
i{\mathcal G}_0^A(t-t^\prime) & = & \theta(t^\prime -t)
[i{\mathcal G}_0^<(t,t^\prime) - i{\mathcal G}_0^>(t,t^\prime)] ,\nonumber
\eeq
with
\begin{eqnarray}  i{\mathcal G}_0^{<}(t,t^\prime) & = &
\frac{{\rm Tr}\{\hat{a}^\dagger(t^\prime)\hat{a}(t)
\hat{\rho}_0\}}{Z}, \nonumber \\
 i{\mathcal G}_0^{>}(t,t^\prime) & = & \frac{{\rm Tr}\{\hat{a}(t)\hat{a}^\dagger(t^\prime)\hat{\rho}_0\}}{Z} . \nonumber
\end{eqnarray}
These expressions can be evaluated using
the interaction representation
\begin{eqnarray}
\hat{a}^\dagger(t^\prime) & = & e^{i(\hat{H}_U - \mu \hat{n})t^\prime}\hat{a}^\dagger e^{-i(\hat{H}_U - \mu\hat{n})t^\prime} \nonumber \\
\hat{a}(t) & = & e^{i(\hat{H}_U -\mu \hat{n})t}\hat{a}e^{-i(\hat{H}_U -\mu \hat{n})t} \nonumber
\end{eqnarray}
we obtain
\begin{eqnarray}
i{\mathcal G}_0^{<}(t,t^\prime) & = & r e^{-i(E_r - \mu r)t}
e^{i(E_{r-1} - \mu(r-1))(t-t^\prime)}e^{i(E_r - \mu r)t^\prime},
\nonumber
\end{eqnarray}
where we recalled $\hat{a}\ket{r} = \sqrt{r}\ket{r-1}$ and
$\hat{a}^\dagger \ket{r} = \sqrt{r+1}\ket{r+1}$.
At temperature $T$,
\begin{eqnarray}
i{\mathcal G}_0^<(t,t^\prime) & = & \frac{\sum_{r=0}^\infty r
e^{i(\mu - U(r-1))(t - t^\prime)}e^{-\beta(E_r - \mu r)}}{
\sum_{r=0}^\infty e^{-\beta(E_r - \mu r)}} . \nonumber \\ & & 
\end{eqnarray}

\begin{widetext}
Hence we have that the retarded Green's function takes the form

\begin{eqnarray}
{\mathcal G}_0^R(t_1,t_2) = -i\theta(t_1 - t_2) \frac{1}{\sum_{r=0}^\infty
e^{-\frac{1}{T}(E_r - \mu r)}}\left\{\sum_{r=0}^\infty
\left[ (r+1)e^{i(\mu - Ur)(t_1 - t_2)} - re^{i(\mu - U(r-1))(t_1 - t_2)}\right]
e^{-\frac{(E_r - \mu r)}{T}}\right\},
\label{eq:retfiniteT}
\end{eqnarray}
which simplifies at $T=0$ to

\begin{eqnarray}
{\mathcal G}_0^R(t_1,t_2) = -i\theta(t_1 - t_2)
\left[(n_0 + 1) e^{i(\mu - Un_0)(t_1 - t_2)} - n_0 e^{i(\mu - U(n_0 -1))(t_1 - t_2)}
\right] .
\label{eq:retzeroT}
\end{eqnarray}
For future reference it will also be convenient to note that
\begin{eqnarray}
{\mathcal G}_0^K(t_1,t_2) = -\frac{i}{\sum_{r=0}^\infty e^{-\beta(E_r - \mu r)}}
\sum_{r=0}^\infty e^{-\beta (E_r - \mu r)} \left[ (r+1) e^{i(\mu - Ur)(t_1 - t_2)} + r e^{i(\mu - U(r-1))(t_1 - t_2)}
\right],
\end{eqnarray}
which simplifies at $T=0$ to
$$ {\mathcal G}_0^K(t_1,t_2) = -i\left[(n_0+1)e^{i(\mu - Un_0)(t_1 - t_2)} + n_0 e^{i(\mu - U(n_0-1))(t_1 - t_2)}\right].$$

Recalling that we can treat this as a single site problem we have
 $${\hat \rho}_0 = e^{-\beta\left[\frac{U}{2}(\hat{n}(\hat{n}-1)-\mu\hat{n})\right]}; \quad Z = 
{\rm Tr}\{\hat{\rho}_0\} = \sum_{r=0}^\infty e^{-\beta(E_r - \mu r)} ,$$
and
 $E_r = \frac{U}{2}r(r-1) $.
$n = N/M$, where $N$ is the number of bosons and $M$ the number of sitesa
 $\mu$ is determined implicitly from
$$ n = \frac{\sum_{r=0}^\infty r e^{-\beta(E_r - \mu r)}}{\sum_{r=0}^\infty e^{-\beta(E_r - \mu r)}}.$$

At $T=0$, the value of $\mu/U$ sets the occupation number,
$n_0\left(\frac{\mu}{U}\right)$ which takes an integer value $r$ for
$r-1 < \mu/U < r$, with degeneracies at $\mu/U = 0, 1, 2, \ldots $.

When we Fourier transform the quadratic part of $S_{\rm eff}^I$ in space and time 
we get:

\begin{eqnarray}
-\int_{-\infty}^\infty \frac{d\omega}{2\pi} \sum_k \frac{1}{J_{\bvec{k}}} \psi_a^*(\omega,\bvec{k})
\tau^1_{ab} \psi_b(\omega,\bvec{k}) 
- \int \frac{d\omega}{2\pi} \sum_{\bvec{k}} 
\psi_{a_1}^*(\omega,\bvec{k}) \tau^1_{a_1 b_1} G_{b_1 b_2}(\omega) \tau^1_{b_2 a_2}
\psi_{a_2}(\omega,\bvec{k}) .
\end{eqnarray}

We choose the hopping amplitude $J_{ij}$ to take the form
$$ J_{ij}(t) = \left\{ \begin{array}{cc} J_0 + j(t), & i,j \, {\rm nearest \,
neighbours} \\
0, & {\rm otherwise} \end{array} \right. , $$
for which (with $a$ the lattice spacing)
\beq
J_{\bvec k}(t) & = & \left[J_0 + j(t)\right]\sum_{j=1}^d \cos(k_ja) \nonumber \\
& \simeq & \left(d   - \frac{1}{2}k^2a^2\right)  \left[J_0 + j(t)\right],
\nonumber
\eeq
assuming that $ka \ll 1$.

Setting $j(t) = 0$ for now, when we take the $\omega$, $k$ $\to$ 0 limit
we can locate the phase boundary by noting when the coefficient of the
$\psi_q^* \psi_c$ term in the
action vanishes:
$$ \frac{1}{2dJ_0} + {\mathcal G}_0^R(\omega = 0) = 0.$$

Note that the retarded propagator
\begin{eqnarray}
{\mathcal G}^R_0(\omega) = \frac{1}{\sum_{r=0}^\infty e^{-\frac{1}{T}(E_r - \mu r)}} \sum_{r=0}^\infty e^{-\frac{(E_r - \mu r)}{T}}
\left[ \frac{(r+1)}{\mu - Ur + \omega + i0} - \frac{r}{\mu - U(r-1) + \omega + i0}\right] ,
\end{eqnarray}
\end{widetext}
at finite $T$ and for $T=0$
\begin{eqnarray}
{\mathcal G}_0^R(\omega) = \frac{n_0+1}{\mu - Un_0 + \omega + i0} - \frac{n_0}{\mu - U(n_0 -1) + \omega + i0} . \nonumber \\
\end{eqnarray}
The advanced propagator may be obtained from
$$ {\mathcal G}_0^A(\omega) = \left[{\mathcal G}_0^R(\omega)\right]^*,$$
and at $T = 0$ the Keldysh propagator is
\begin{eqnarray}
{\mathcal G}_0^K(\omega) & = & - 2i\pi \left[(n_0+1) 
\delta(\omega + \mu - Un_0) 
\right. \nonumber \\ & & \left. +  
n_0 \delta(\omega + \mu - U(n_0 - 1))\right] 
\end{eqnarray}
At zero temperature we obtain the standard mean field equation for the 
phase boundary between the Mott insulator and superfluid phases:
$$ \frac{1}{2dJ_0} + \frac{(n_0 + 1)}{\mu - Un_0} - \frac{n_0}{\mu - U(n_0-1)} = 0.$$
This may also be expressed as

\begin{eqnarray}
\tilde{\mu}_\pm = \frac{1}{2}\left[(2n_0 + 1) - \tilde{J} \pm
\sqrt{1 - \tilde{J}(2n_0 + 1) + \tilde{J}^2}\right], \nonumber \\
\end{eqnarray}
for $n_0 > 1$ and $\tilde{\mu}_+ = - \tilde{J}$ if $n_0 = 0$,
where $\tilde{J} = 2dJ/U$ and $\tilde{\mu} = \mu/U$.
 This affirms that the effective action 
correctly predicts the mean field phase boundary at zero temperature. At finite
temperature the corresponding equation is
\begin{eqnarray} 
\frac{1}{2dJ_0} + \frac{1}{Z}\sum_{r=0}^\infty e^{-\beta(E_r - \mu r)}
\left[\frac{r+1}{\mu - Ur} - \frac{r}{\mu - U(r-1)}\right] = 0 . \nonumber  \\
\label{eq:boundary}
\end{eqnarray}
The phase boundary as determined from this equation for a variety of $\beta$
values is displayed in Fig.~\ref{fig:parameters} d).  
This phase boundary is the edge of the 
superfluid phase -- the Mott insulator is strictly defined only at $T=0$, and
at non-zero temperature there can be a
normal phase separating superfluid and insulator, with
full melting of the insulator for $T^* \simeq 0.2 \, U$ \cite{Mahmud,Gerbier}.

\begin{widetext}
\section{Parameters in the equation of motion}
\label{app:eqmot}
There are three parameters that enter the equation of motion:
$$ \nu = \left[{\mathcal G}_0^R\right]^{-1}_{\omega= 0}; \lambda = -\left.\frac{\partial}{\partial\omega}
\left[{\mathcal G}_0^R\right]^{-1} \right|_{\omega = 0}; 
\kappa^2 = \left. \frac{1}{2}\frac{\partial^2}{\partial \omega^2} \left[{\mathcal G}_0^R\right]^{-1}
\right|_{\omega=0}.$$ These can be evaluated to give

\begin{eqnarray}
\nu = \frac{Z}{\sum_{r=0}^\infty e^{-\beta(E_r - \mu r)} \left[ \frac{(r+1)}{\mu - Ur} - \frac{r}{\mu - U(r-1)}
\right]}, 
\end{eqnarray}
\begin{eqnarray}
\lambda = \frac{\nu^2}{Z} 
\sum_{r=0}^\infty e^{-\beta(E_r - \mu r)} \left[\frac{(r+1)}{[\mu - Ur]^2} - \frac{r}{[\mu - U(r-1)]^2}
\right]  ,
\end{eqnarray}
\begin{eqnarray}
\kappa^2 & = & \frac{\lambda^2}{\nu} - \frac{\nu^2}{Z}
\sum_{r=0}^\infty e^{-\beta(E_r - \mu r)} \left[\frac{(r+1)}{[\mu - Ur]^3} - 
\frac{r}{[\mu - U(r-1)]^3}\right],
\end{eqnarray}
and
\begin{eqnarray}
u & = & - \frac{\nu^4}{2Z} \sum_{r=0}^\infty e^{-\beta(E_r - \mu_r)} \left\{ 
\frac{4(p+1)(p+2)}{[Up - \mu]^2[2\mu - (2p+1)U]} 
 + \frac{4p(p-1)}{[U(p-1)-\mu]^2[U(2p-3) - 2\mu]} 
 - \frac{4(p+1)^2}{[\mu - Up]^3} \right. \nonumber \\
& & \hspace*{3.5cm} \left.
 - \frac{4p^2}{[U(p-1)-\mu]^3}
 - \frac{4p(p+1)}{[U(p-1)-\mu]^2[\mu - Up]}
 - \frac{4p(p+1)}{[U(p-1)-\mu][\mu - Up]^2} \right\} .
\end{eqnarray}
The expressions for $\nu$, $\lambda$ and $\kappa^2$ simplify somewhat in the zero temperature limit:
$$\nu = \frac{(\mu - Un_0)(\mu - U(n_0 - 1))}{\mu + U}; \quad \quad 
\lambda = \frac{(2n_0 -1)U - 2\mu}{\mu + U} + \frac{(\mu -Un_0)(\mu - U(n_0 -1))}{(\mu + U)^2}
$$
and
\begin{eqnarray}
\kappa^2 = \frac{1}{2}\left[\frac{2n_0 U - \mu}{(\mu + U)^2} 
 - \left\{\frac{(U\mu(2n_0 +1) - U^2(2n_0^2 - 1)}{(\mu + U)^3}\right\} \right].
\end{eqnarray}

\section{Evaluation of the four point function}
\label{app:fourpoint}
To evaluate the four time correlation functions, there are several basic
correlations we need:

\begin{eqnarray}
B^{aa a^\dagger a^\dagger}(t_1,t_2,t_3,t_4) & = & \frac{1}{Z} {\rm Tr} 
\left\{ e^{-\beta(\hat{H}_U - \mu \hat{N})} a(t_1) a(t_2) a^\dagger(t_3)
a^\dagger(t_4)\right\}  \\
& = & \frac{1}{Z} \sum_{p=0}^\infty
(p+1)(p+2) e^{i(E_p - \mu p)(t_1 - t_4 + i\beta) + 
i(E_{p+1} - \mu (p+1))(t_2 + t_4 - t_1 - t_3) 
+ i(E_{p+2} - \mu (p+2))(t_3 - t_2)} , \nonumber \\
B^{aa^\dagger a a^\dagger}(t_1,t_2,t_3,t_4) & = & \frac{1}{Z} \sum_{p=0}^\infty
(p+1)^2 e^{i(E_p - \mu p)(t_1 + t_3 - t_2 -t_4 + i\beta) 
+ i(E_{p+1} -\mu (p+1))(t_2 + t_4 - t_1 - t_3)} ,  \\
B^{aa^\dagger a^\dagger a}(t_1,t_2,t_3,t_4) & = & \frac{1}{Z} \sum_{p=0}^\infty
p(p+1) e^{i(E_p - \mu p)(t_1 + t_3 - t_2 -t_4 + i\beta) + i(E_{p-1} - \mu(p-1))
(t_4 - t_3) + i(E_{p+1} - \mu(p+1))(t_2 - t_1)} ,  \\
B^{a^\dagger a a a^\dagger}(t_1,t_2,t_3,t_4) & = & \frac{1}{Z} \sum_{p=0}^\infty
p(p+1) e^{i(E_p - \mu p)(t_1 + t_3 - t_2 -t_4 + i\beta) + i(E_{p+1} - \mu(p+1))
(t_4 - t_3) + i(E_{p-1} - \mu(p-1))(t_2 - t_1)} ,  \\
B^{a^\dagger a a^\dagger a}(t_1,t_2,t_3,t_4) & = & \frac{1}{Z} \sum_{p=0}^\infty
p^2 e^{i(E_p - \mu p)(t_1 + t_3 - t_2 -t_4 + i\beta)
+ i(E_{p-1} -\mu (p-1))(t_2 + t_4 - t_1 - t_3)} ,  \\
B^{a^\dagger a^\dagger a a}(t_1,t_2,t_3,t_4) & = & \frac{1}{Z} \sum_{p=0}^\infty
p(p-1) e^{i(E_p - \mu p)(t_1 - t_4 + i\beta) +
i(E_{p-1} - \mu (p-1))(t_2 + t_4 - t_1 - t_3)
+ i(E_{p-2} - \mu (p-2))(t_3 - t_2)} .
\end{eqnarray}

In addition we require the two point correlations

\begin{eqnarray}
C^{a a^\dagger}(t_1,t_2) & = & \frac{1}{Z} {\rm Tr}\left\{
e^{-\beta(\hat{H}_U - \mu \hat{N})} a(t_1) a^\dagger(t_2)\right\} \nonumber \\
& = & \frac{1}{Z} \sum_{p=0}^\infty (p+1) 
e^{i(E_p - \mu p)(t_1 - t_2 + i\beta) + i(E_{p+1} - \mu(p+1))(t_2-t_1)} \nonumber \\
& = & i{\mathcal G}_0^>(t_1,t_2), \\
C^{a^\dagger a}(t_1,t_2) & = & \frac{1}{Z} \sum_{p=0}^\infty p
e^{i(E_p - \mu p)(t_1 - t_2 + i\beta) + i(E_{p-1} - \mu(p-1))(t_2-t_1)} \nonumber \\
& = & i{\mathcal G}_0^<(t_2,t_1) . 
\end{eqnarray}

The actual expressions are rather tiresome to derive but are given here
for completeness, where we use the notation $\theta_{ij} = \theta(t_i - t_j)$:

\begin{eqnarray}
G^{2c}_{qqqq}(t_1,t_2,t_3,t_4) & = & \frac{i}{2} \left\{ 
\left[\theta_{12}\theta_{23} + \theta_{21} \theta_{14}\right]
\left(B^{a^\dagger aa a^\dagger}(t_3,t_2,t_1,t_4) 
+ B^{a^\dagger aa a^\dagger}(t_4,t_1,t_2,t_3) \right) \right. \nonumber \\
& & + \left[\theta_{12}\theta_{24} + \theta_{21} \theta_{13}\right]
\left(B^{a^\dagger aa a^\dagger}(t_3,t_1,t_2,t_4) 
+ B^{a^\dagger aa a^\dagger}(t_4,t_2,t_1,t_3) \right) \nonumber \\
& & + \left[\theta_{13}\theta_{32} + \theta_{31} \theta_{14}\right]
\left(B^{a a^\dagger a a^\dagger}(t_2,t_3,t_1,t_4) 
+ B^{a^\dagger a a^\dagger a}(t_4,t_1,t_3,t_2) \right) \nonumber \\
& & + \left[\theta_{13} \theta_{34} + \theta_{31} \theta_{12}\right]
\left(B^{aa a^\dagger  a^\dagger}(t_2,t_1,t_3,t_4) 
+ B^{a^\dagger aa a^\dagger}(t_4,t_3,t_1,t_2) \right) \nonumber \\
& & + \left[\theta_{14}\theta_{42} + \theta_{41} \theta_{13}\right]
\left(B^{ a a^\dagger a a^\dagger}(t_2,t_4,t_1,t_3) 
+ B^{a^\dagger a a^\dagger a}(t_3,t_1,t_4,t_2) \right) \nonumber \\
&  & + \left[\theta_{14}\theta_{43} + \theta_{41} \theta_{12}\right]
\left(B^{ aa a^\dagger  a^\dagger}(t_2,t_1,t_4,t_3)
+ B^{a^\dagger a^\dagger a a}(t_3,t_4,t_1,t_2) \right) \nonumber \\
& & +  \left[\theta_{32}\theta_{21} + \theta_{23} \theta_{34}\right]
\left(B^{ aa a^\dagger  a^\dagger}(t_1,t_2,t_3,t_4)
+ B^{a^\dagger a^\dagger a a}(t_4,t_3,t_2,t_1) \right) \nonumber \\
& & +  \left[\theta_{42}\theta_{21} + \theta_{24} \theta_{43}\right]
\left(B^{ aa a^\dagger  a^\dagger}(t_1,t_2,t_4,t_3)
+ B^{a^\dagger a^\dagger a a}(t_3,t_4,t_2,t_1) \right) \nonumber \\
& & +  \left[\theta_{23}\theta_{31} + \theta_{32} \theta_{24}\right]
\left(B^{ a a^\dagger a  a^\dagger}(t_1,t_3,t_2,t_4)
+ B^{a^\dagger a a^\dagger  a}(t_4,t_2,t_3,t_1) \right) \nonumber \\
& & +  \left[\theta_{43} \theta_{31} + \theta_{34} \theta_{42}\right]
\left(B^{ a a^\dagger a^\dagger a}(t_1,t_3,t_4,t_2)
+ B^{a a^\dagger a^\dagger  a}(t_2,t_4,t_3,t_1) \right) \nonumber \\
& & +  \left[\theta_{24}\theta_{41} + \theta_{42} \theta_{23}\right]
\left(B^{ a a^\dagger a a^\dagger }(t_1,t_4,t_2,t_3)
+ B^{a^\dagger a a^\dagger  a}(t_3,t_2,t_4,t_1) \right) \nonumber \\
& & \left. +  \left[\theta_{34}\theta_{41} + \theta_{43} \theta_{32}\right]
\left(B^{ a a^\dagger a^\dagger a}(t_1,t_4,t_3,t_2)
+ B^{a a^\dagger a^\dagger  a}(t_2,t_3,t_4,t_1) \right) \right\} \nonumber \\
& & -i\left\{ \left[C^{aa^\dagger}(t_1,t_3) + C^{a^\dagger a}(t_3,t_1)\right]
\left[C^{aa^\dagger}(t_2,t_4) + C^{a^\dagger a}(t_4,t_2)\right] \right.
\nonumber \\
& & \left. +  \left[C^{aa^\dagger}(t_1,t_4) + C^{a^\dagger a}(t_4,t_1)\right]
\left[C^{aa^\dagger}(t_2,t_3) + C^{a^\dagger a}(t_3,t_2)\right] \right\} ,
\label{eq:Gqqqq}
\end{eqnarray}

\begin{eqnarray}
G^{2c}_{cqqq}(t_1,t_2,t_3,t_4)& = & \frac{i}{2}\left\{ -\theta_{21}
\left[\theta_{32} + \theta_{23}\theta_{34}\right]\left(
B^{aaa^\dagger a^\dagger}(t_1,t_2,t_3,t_4) -
B^{a^\dagger a^\dagger a a}(t_4,t_3,t_2,t_1) \right) \right. \nonumber \\
& & -  \theta_{21}
\left[\theta_{42} + \theta_{24}\theta_{43}\right]
\left(
B^{aaa^\dagger a^\dagger}(t_1,t_2,t_4,t_3) -
B^{a^\dagger a^\dagger a a}(t_3,t_4,t_2,t_1) \right) \nonumber \\
& & - \theta_{31}
\left[\theta_{23} + \theta_{32}\theta_{24}\right]
\left( B^{aa^\dagger aa^\dagger}(t_1,t_3,t_2,t_4)  -
B^{a^\dagger a a^\dagger  a}(t_4,t_2,t_3,t_1) \right) \nonumber \\
& & - \theta_{31}
\left[\theta_{43} + \theta_{34}\theta_{42}\right]
\left( B^{aa^\dagger a^\dagger a}(t_1,t_3,t_4,t_2) -
B^{a a^\dagger a^\dagger  a}(t_2,t_4,t_3,t_1) \right) \nonumber \\
& & - \theta_{41}
\left[\theta_{24} + \theta_{42}\theta_{23}\right]
\left( B^{aa^\dagger a a^\dagger }(t_1,t_4,t_2,t_3) -
B^{a^\dagger a a^\dagger  a}(t_3,t_2,t_4,t_1) \right)  \nonumber \\
& & - \theta_{41}
\left[\theta_{34} + \theta_{43}\theta_{32}\right]
\left( B^{aa^\dagger a^\dagger a}(t_1,t_4,t_3,t_2) -
B^{a a^\dagger a^\dagger  a}(t_2,t_3,t_4,t_1) \right)  \nonumber \\
& & -  \left[\theta_{21}\theta_{13}
- \theta_{31}\theta_{12} \theta_{24}\right]
\left(B^{a^\dagger aa a^\dagger }(t_3,t_1,t_2,t_4) -
B^{a^\dagger a a a^\dagger}(t_4,t_2,t_1,t_3) \right)  \nonumber \\
& & +  \left[\theta_{21}\theta_{14} - 
\theta_{41}\theta_{12} \theta_{23}\right]
\left(B^{a^\dagger a a a^\dagger}(t_3,t_2,t_1,t_4) -
B^{a^\dagger aa a^\dagger }(t_4,t_1,t_2,t_3)\right)  \nonumber \\
& & - \left[\theta_{31}\theta_{12} - 
\theta_{21}\theta_{13} \theta_{34}\right] 
\left(B^{a a a^\dagger a^\dagger }(t_2,t_1,t_3,t_4) -
B^{a^\dagger a^\dagger a a}(t_4,t_3,t_1,t_2) \right)  \nonumber \\
& & + \left[\theta_{31}\theta_{14} -
\theta_{41}\theta_{13}\theta_{32} \right]
\left(B^{a a^\dagger a a^\dagger}(t_2,t_3,t_1,t_4) 
- B^{a^\dagger a a^\dagger a}(t_4,t_1,t_3,t_2)\right)  \nonumber \\
& & - \left[\theta_{41}\theta_{12} -
\theta_{21}\theta_{14}\theta_{43} \right]
\left(
B^{aa a^\dagger a^\dagger }(t_2,t_1,t_4,t_3) -
B^{a^\dagger a^\dagger aa}(t_3,t_4,t_1,t_2) \right)  \nonumber \\
& & \left. + \left[\theta_{41}\theta_{13} -
\theta_{31}\theta_{14}\theta_{42} \right]
\left(B^{a a^\dagger a a^\dagger}(t_2,t_4,t_1,t_3) - 
B^{a^\dagger a a^\dagger a}(t_3,t_1,t_4,t_2)\right)  \right\} \nonumber \\
& & -i \left\{ \theta_{31} \left[C^{a^\dagger a}(t_3,t_1) - 
C^{aa^\dagger}(t_1,t_3)\right]
\left[C^{aa^\dagger}(t_2,t_4) + C^{a^\dagger a}(t_4,t_2)\right] \right. 
\nonumber \\
& & \left. + \theta_{41} \left[C^{a^\dagger a}(t_4,t_1) - 
C^{aa^\dagger}(t_1,t_4)\right]
\left[C^{aa^\dagger}(t_2,t_3) + C^{a^\dagger a}(t_3,t_2)\right] \right\} ,
\label{eq:Gcqqq}
\end{eqnarray}

\begin{eqnarray}
G^{2c}_{ccqq}(t_1,t_2,t_3,t_4) & = & \frac{i}{2} \left\{ \theta_{32} \theta_{21}
\left( B^{aaa^\dagger a^\dagger}(t_1,t_2,t_3,t_4) + B^{a^\dagger a^\dagger aa}(t_4,t_3,t_2,t_1)\right) \right. \nonumber \\
& &  +  \theta_{42} \theta_{21} \left(
B^{aa a^\dagger a^\dagger}(t_1,t_2,t_4,t_3) + B^{a^\dagger a^\dagger aa}(t_3,t_4,t_2,t_1)\right) \nonumber \\
& & + \theta_{31}\left[\theta_{42}\theta_{23} - \theta_{32} \theta_{24}\right] 
\left( B^{aa^\dagger a a^\dagger}(t_1,t_3,t_2,t_4)+ B^{a^\dagger a a^\dagger a}(t_4, t_2, t_3, t_1) \right)
\nonumber \\
& & + \theta_{41} \left[\theta_{32}\theta_{24} - \theta_{42} \theta_{23}) \right] 
\left( B^{aa^\dagger a a^\dagger}(t_1,t_4,t_2,t_3) + B^{a^\dagger a a^\dagger a}(t_3,t_2,t_4,t_1)\right) \nonumber \\
& & -\theta_{31} \theta_{42} \left(
B^{aa^\dagger a^\dagger a}(t_1,t_3,t_4,t_2) +
B^{a a^\dagger a^\dagger a}(t_2,t_4,t_3,t_1) \right) \nonumber \\
& & - \theta_{41}\theta_{32} \left(
B^{a a^\dagger a^\dagger a}(t_1,t_4,t_3,t_2) 
+   B^{a a^\dagger a^\dagger a}(t_2,t_3,t_4,t_1)\right)
 \nonumber \\
& & + \theta_{31}\theta_{12} \left( B^{aaa^\dagger a^\dagger}(t_2,t_1,t_3,t_4)
+ B^{a^\dagger a^\dagger a a}(t_4,t_3,t_1,t_2) \right) \nonumber \\
& & + \theta_{41} \theta_{12}\left(  B^{aa a^\dagger a^\dagger}(t_2,t_1,t_4,t_3)
+ B^{a^\dagger a^\dagger aa}(t_3,t_4,t_1,t_2) \right)
 \nonumber \\
& & + \theta_{32} \left[\theta_{41}\theta_{13} - \theta_{31} \theta_{14}\right] 
\left( B^{aa^\dagger a a^\dagger}(t_2,t_3,t_1,t_4) +  B^{a^\dagger a a^\dagger a}(t_4,t_1,t_3,t_2) \right) \nonumber \\
& & + \theta_{42} \left[\theta_{31} \theta_{14} - \theta_{41} \theta_{13}\right]
\left( B^{a a^\dagger a a^\dagger}(t_2,t_4,t_1,t_3) + B^{a^\dagger aa^\dagger a}(t_3,t_1,t_4,t_2) \right) \nonumber \\
& & + \left[\theta_{31} \theta_{12} \theta_{24} + \theta_{42} \theta_{21}
 \theta_{13} \right] \left(B^{a^\dagger aa a^\dagger}(t_3,t_1,t_2,t_4) 
+ B^{a^\dagger a a a^\dagger}(t_4,t_2,t_1,t_3) \right) \nonumber \\
& & + \left.\left[\theta_{41} \theta_{12} \theta_{23} + 
\theta_{32} \theta_{21} \theta_{14}\right] \left( 
B^{a^\dagger aa a^\dagger}(t_3,t_2,t_1,t_4) + 
B^{a^\dagger aaa^\dagger}(t_4,t_1,t_2,t_3) \right)
\right\} \nonumber \\
& & -i \left\{ \theta_{31}\theta_{42}
 \left(C^{aa^\dagger}(t_2,t_4) - 
C^{a^\dagger a}(t_4,t_2) \right)\left(C^{a^\dagger a}(t_3,t_1) - C^{aa^\dagger}(t_1,t_3)
\right) \right. \nonumber \\
& & +\left. \theta_{41} \theta_{32}  \left(C^{aa^\dagger}(t_2,t_3) -
C^{a^\dagger a}(t_3,t_2) \right)\left(C^{a^\dagger a}(t_4,t_1) - C^{aa^\dagger}(t_1,t_4)\right) \right\} ,
\end{eqnarray}

\begin{eqnarray}
G^{2c}_{cqcq}(t_1,t_2,t_3,t_4) & = & \frac{i}{2} \left\{ \theta_{21}
\left[\theta_{43} \theta_{32} - \theta_{23} \theta_{34} 
\right] \left(B^{aaa^\dagger a^\dagger}(t_1,t_2,t_3,t_4)
+ B^{a^\dagger a^\dagger aa}(t_4,t_3,t_2,t_1)\right) \right. \nonumber \\
& & +  \theta_{41}
\left[\theta_{23} \theta_{34} - \theta_{43} \theta_{32}
\right] \left( B^{aaa^\dagger a^\dagger}(t_1,t_4,t_3,t_2)
+ B^{aa^\dagger a^\dagger a}(t_2,t_3,t_4,t_1)\right)
 \nonumber \\
& & + \theta_{43}\left[\theta_{21}\theta_{14} - \theta_{41}
\theta_{12}\right]\left(B^{aaa^\dagger a^\dagger}(t_2,t_1,t_4,t_3)
+ B^{a^\dagger a^\dagger aa}(t_3,t_4,t_1,t_2)\right) \nonumber \\
& & + \theta_{23}\left[\theta_{41}\theta_{12} - \theta_{21} \theta_{14}\right]\left(B^{a^\dagger aa a^\dagger}(t_3,t_2,t_1,t_4)
+  B^{a^\dagger a a a^\dagger}(t_4,t_1,t_2,t_3) \right) \nonumber \\
& & - \theta_{21}\theta_{43}
\left( B^{aaa^\dagger a^\dagger}(t_1,t_2,t_4,t_3) 
+ B^{a^\dagger a^\dagger aa}(t_3,t_4,t_2,t_1) \right)
 \nonumber \\
& & + \theta_{23} \theta_{31} \left(
B^{aa^\dagger aa^\dagger}(t_1,t_3,t_2,t_4) + B^{a^\dagger a a^\dagger a}(t_4,t_2,t_3,t_1)
\right)
\nonumber \\
& & + \theta_{43} \theta_{31} \left( 
B^{aa^\dagger a^\dagger a}(t_1,t_3,t_4,t_2) 
+ B^{aa^\dagger a^\dagger a}(t_2,t_4,t_3,t_1) \right)
\nonumber \\
& & - \theta_{23} \theta_{41} \left(
 B^{aa^\dagger aa^\dagger}(t_1,t_4,t_2,t_3) +  B^{a^\dagger a a^\dagger a}(t_3,t_2,t_4,t_1)
\right)
\nonumber \\
& & + \theta_{41} \theta_{13} \left(
B^{aa^\dagger a a^\dagger}(t_2,t_4,t_1,t_3) + B^{a^\dagger a a^\dagger a}(t_3,t_1,t_4,t_2)
\right)
\nonumber \\
& & + \theta_{21} \theta_{13} \left(
B^{a^\dagger a a a^\dagger}(t_3,t_1,t_2,t_4)
+ B^{a^\dagger a a a^\dagger}(t_4,t_2,t_1,t_3) \right)
\nonumber \\
& & + \left[\theta_{21}\theta_{13}\theta_{34} +
\theta_{43}\theta_{31} \theta_{12} \right] \left(
B^{aaa^\dagger a^\dagger}(t_2,t_1,t_3,t_4)
+ B^{a^\dagger a^\dagger aa}(t_4,t_3,t_1,t_2) \right)
 \nonumber \\
& & + \left. \left[\theta_{23}\theta_{31}\theta_{14} +
\theta_{41}\theta_{13} \theta_{32} \right] \left(
B^{aa^\dagger a a^\dagger}(t_2,t_3,t_1,t_4)
+ B^{a^\dagger a a^\dagger a}(t_4,t_1,t_3,t_2)
\right)  \right\} \nonumber \\
& & - i \theta_{23} \theta_{41} \left[C^{a^\dagger a}(t_4,t_1) -
C^{aa^\dagger}(t_1,t_4)\right]
\left[C^{aa^\dagger}(t_2,t_3) - C^{a^\dagger a}(t_3,t_2)
\right]  ,
\end{eqnarray}

\begin{eqnarray}
G^{2c}_{cccq}(t_1,t_2,t_3,t_4) & = & \frac{i}{2} \left\{ 
\theta_{43}\theta_{32} \theta_{21} \left(
B^{a^\dagger a^\dagger aa}(t_4,t_3,t_2,t_1) -
B^{aaa^\dagger a^\dagger}(t_1,t_2,t_3,t_4) \right)
 \right. \nonumber \\
& & + \theta_{43}\theta_{42} \theta_{21} \left(
B^{aaa^\dagger a^\dagger}(t_1,t_2,t_4,t_3)
- B^{a^\dagger a^\dagger aa}(t_3,t_4,t_2,t_1) \right)
  \nonumber \\
& & - \theta_{42}\theta_{23} \theta_{31} 
\left( B^{aa^\dagger a a^\dagger}(t_1,t_3,t_2,t_4) -
B^{a^\dagger a a^\dagger a}(t_4,t_2,t_3,t_1) \right)
 \nonumber \\
& & + \theta_{43}\theta_{31} \theta_{42} \left(
B^{aa^\dagger a^\dagger a}(t_1,t_3,t_4,t_2)
- B^{aa^\dagger a^\dagger a}(t_2,t_4,t_3,t_1)  \right)  \nonumber \\
& & - \theta_{41}\theta_{42} \theta_{23} \left(
B^{aa^\dagger a a^\dagger}(t_1,t_4,t_2,t_3)
- B^{a^\dagger a a^\dagger a}(t_3,t_2,t_4,t_1) \right)  \nonumber \\
& & - \theta_{43}\theta_{32} \theta_{41} \left(
B^{aa^\dagger a^\dagger a}(t_1,t_4,t_3,t_2)
- B^{aa^\dagger a^\dagger a}(t_2,t_3,t_4,t_1) \right)  \nonumber \\
& & - \theta_{43}\theta_{31} \theta_{12} \left(
B^{aaa^\dagger a^\dagger}(t_2,t_1,t_3,t_4)  -
B^{a^\dagger a^\dagger aa}(t_4,t_3,t_1,t_2) \right)
  \nonumber \\
& & + \theta_{41}\theta_{12} \theta_{43} \left(
B^{aaa^\dagger a^\dagger}(t_2,t_1,t_4,t_3) 
- B^{a^\dagger a^\dagger aa}(t_3,t_4,t_1,t_2) \right) \nonumber \\
& & + \theta_{41}\theta_{13} \theta_{32} 
\left( B^{aa^\dagger a a^\dagger}(t_2,t_3,t_1,t_4) -
B^{a^\dagger a a^\dagger a}(t_4,t_1,t_3,t_2)  \right)
 \nonumber \\
& & - \theta_{41}\theta_{13} \theta_{42} \left(
B^{aa^\dagger a a^\dagger}(t_2,t_4,t_1,t_3)
 - B^{a^\dagger a a^\dagger a}(t_3,t_1,t_4,t_2) \right)  \nonumber \\
& & - \theta_{42}\theta_{21} \theta_{13} \left(
B^{a^\dagger aa a^\dagger }(t_3,t_1,t_2,t_4) 
- B^{a^\dagger aa a^\dagger}(t_4,t_2,t_1,t_3)
\right) \nonumber \\
& & - \left. \theta_{41}\theta_{12} \theta_{23} \left(
B^{a^\dagger a a a^\dagger}(t_3,t_2,t_1,t_4)  -
B^{a^\dagger aa a^\dagger}(t_4,t_1,t_2,t_3)  \right)
  \right\} ,
\end{eqnarray}

\begin{eqnarray}
G^{2c}_{qqqc}(t_1,t_2,t_3,t_4) & = & \frac{i}{2}\left\{\theta_{34}\left[\theta_{23} + \theta_{32}\theta_{21}\right]
\left(B^{aaa^\dagger a^\dagger}(t_1,t_2,t_3,t_4) - B^{a^\dagger a^\dagger aa}(t_4,t_3,t_2,t_1)\right) \right. \nonumber \\
& & + \left[\theta_{24}\theta_{43} - \theta_{34} \theta_{42}\theta_{21}\right] 
\left(B^{aaa^\dagger a^\dagger}(t_1,t_2,t_4,t_3) - B^{a^\dagger a^\dagger a a}(t_3,t_4,t_2,t_1)\right) \nonumber \\
& & +\theta_{24}\left[\theta_{32} + \theta_{23}\theta_{31}\right] 
\left(B^{aa^\dagger a a^\dagger}(t_1,t_3,t_2,t_4) - B^{a^\dagger aa^\dagger a}(t_4,t_2,t_3,t_1)\right) \nonumber \\
& & +\left[\theta_{34}\theta_{42} - \theta_{24}\theta_{43}\theta_{31}\right]
\left(B^{aa^\dagger a^\dagger a}(t_1,t_3,t_4,t_2) - B^{aa^\dagger a^\dagger a}(t_2,t_4,t_3,t_1)\right) \nonumber \\
& & +\left[\theta_{14}\theta_{42}\theta_{23} - \theta_{24}\theta_{41} \right]
\left(B^{aa^\dagger aa^\dagger}(t_1,t_4,t_2,t_3) - B^{a^\dagger a a^\dagger a}(t_3,t_2,t_4,t_1)\right) \nonumber \\
& & +\left[\theta_{14}\theta_{43}\theta_{32} - \theta_{34}\theta_{41}\right]
\left(B^{aa^\dagger a^\dagger a}(t_1,t_4,t_3,t_2) - B^{aa^\dagger a^\dagger a}(t_2,t_3,t_4,_1)\right) \nonumber \\
& & +\theta_{34}\left[\theta_{13} + \theta_{31}\theta_{12}\right]
\left(B^{aaa^\dagger a^\dagger}(t_2,t_1,t_3,t_4) - B^{a^\dagger a^\dagger aa}(t_4,t_3,t_1,t_2)\right) \nonumber \\
& & +\left[\theta_{14}\theta_{43} -\theta_{34}\theta_{41}\theta_{12}\right]
\left(B^{aaa^\dagger a^\dagger}(t_2,t_1,t_4,t_3) - B^{a^\dagger a^\dagger aa}(t_3,t_4,t_1,t_2)\right) \nonumber \\
& & +\theta_{14}\left[\theta_{31} + \theta_{13}\theta_{32} \right] 
\left(B^{aa^\dagger a a^\dagger}(t_2,t_3,t_1,t_4) - B^{a^\dagger a a^\dagger a}(t_4,t_1,t_3,t_2)\right) \nonumber \\
& & + \left[\theta_{24} \theta_{41}\theta_{13} - \theta_{14}\theta_{42}\right]
\left(B^{aa^\dagger a a^\dagger}(t_2,t_4,t_1,t_3) - B^{a^\dagger a a^\dagger a}(t_3,t_1,t_4,t_2) \right) \nonumber \\
& & + \theta_{24} \left[\theta_{12} + \theta_{21}\theta_{13}\right]
\left(B^{a^\dagger aaa^\dagger}(t_3,t_1,t_2,t_4) - B^{a^\dagger aaa^\dagger}(t_4,t_2,t_1,t_3)\right) \nonumber \\
& & \left. + \theta_{14} \left[\theta_{21} + \theta_{12}\theta_{23}\right]
\left(B^{a^\dagger aaa^\dagger}(t_3,t_2,t_1,t_4) - B^{a^\dagger a a a^\dagger}(t_4,t_1,t_2,t_3) \right) \right\} \nonumber \\
& & - i \left\{ \theta_{24}\left(C^{aa^\dagger}(t_1,t_3) + C^{a^\dagger a}(t_3,t_1)\right)\left(C^{aa^\dagger}(t_2,t_4) 
 - C^{a^\dagger a}(t_4,t_2)\right) \right. \nonumber \\
& & \left. + \theta_{14} \left(C^{aa^\dagger}(t_2,t_3) + C^{a^\dagger a}(t_3,t_2)\right)
\left(C^{aa^\dagger}(t_1,t_4) - C^{a^\dagger a}(t_4,t_1) \right) \right\} ,
\end{eqnarray}

\begin{eqnarray}
G_{qqcc}^{2c}(t_1,t_2,t_3,t_4) & = & \frac{i}{2}\left\{ \theta_{23} \theta_{34} \left(B^{aaa^\dagger a^\dagger}(t_1,t_2,t_3,t_4)
 + B^{a^\dagger a^\dagger aa}(t_4,t_3,t_2,t_1) \right) \right. \nonumber \\
& & + \theta_{24} \theta_{43} \left(B^{aaa^\dagger a^\dagger}(t_1,t_2,t_4,t_3) + B^{a^\dagger a^\dagger aa}(t_3,t_4,t_1,t_2)\right)
\nonumber \\
& & + \theta_{24}\left[\theta_{13}\theta_{32} - \theta_{23}\theta_{31} \right] 
\left(B^{aa^\dagger aa^\dagger}(t_1,t_3,t_2,t_4)
 + B^{a^\dagger aa^\dagger a}(t_4,t_2,t_3,t_1)\right) \nonumber \\
& & + \left[\theta_{13} \theta_{34} \theta_{42} + \theta_{24} \theta_{43} \theta_{31} \right] 
\left(B^{aa^\dagger a^\dagger a}(t_1,t_3,t_4,t_2) + B^{aa^\dagger a^\dagger a}(t_2,t_4,t_3,t_1)\right) \nonumber \\
& & + \theta_{23} \left[\theta_{14} \theta_{42} - \theta_{24}\theta_{41}\right] 
\left(B^{aa^\dagger aa^\dagger}(t_1,t_4,t_2,t_3) + B^{a^\dagger aa^\dagger a}(t_3,t_2,t_4,t_1)\right) \nonumber \\
& & + \left[\theta_{14}\theta_{43}\theta_{32} + \theta_{23} \theta_{34} \theta_{41} \right]
\left( B^{aa^\dagger aa^\dagger}(t_1,t_4,t_3,t_2) + B^{aa^\dagger a^\dagger a}(t_2,t_3,t_4,t_1)\right) \nonumber \\
& & + \theta_{13} \theta_{34} \left(B^{aaa^\dagger a^\dagger}(t_2,t_1,t_3,t_4) + B^{a^\dagger a^\dagger aa}(t_4,t_3,t_1,t_2)
\right) \nonumber \\
& & +\theta_{14}\theta_{43} \left(B^{aaa^\dagger a^\dagger}(t_2,t_1,t_4,t_3) + B^{a^\dagger a^\dagger aa}(t_3,t_4,t_1,t_2)
\right) \nonumber \\
& & + \theta_{14} \left[\theta_{23}\theta_{31} - \theta_{13}\theta_{32}\right]
\left(B^{aa^\dagger aa^\dagger}(t_2,t_3,t_1,t_4) + B^{a^\dagger a a^\dagger a}(t_4,t_1,t_3,t_2)\right) \nonumber \\
& & + \theta_{13} \left[\theta_{24}\theta_{41} - \theta_{14}\theta_{42}\right] 
\left(B^{aa^\dagger aa^\dagger}(t_2,t_4,t_1,t_3) + B^{a^\dagger a a^\dagger a}(t_3,t_1,t_4,t_2)\right) \nonumber \\
& & - \theta_{13} \theta_{24} \left(B^{a^\dagger aaa^\dagger}(t_3,t_1,t_2,t_4) + B^{a^\dagger aaa^\dagger}(t_4,t_2,t_1,t_3)
\right) \nonumber \\
& & -\left. \theta_{14}\theta_{23} \left(B^{a^\dagger aaa^\dagger}(t_3,t_2,t_1,t_4) + B^{a^\dagger aaa^\dagger}(t_4,t_1,t_2,t_3)\right)
\right\} \nonumber \\
& & + i\left\{ \theta_{13} \theta_{24} \left[C^{a^\dagger a}(t_3,t_1) - C^{aa^\dagger}(t_1,t_3)\right]
\left[C^{a^\dagger a}(t_4,t_2) - C^{aa^\dagger}(t_2,t_4)\right] \right. \nonumber \\
& & \left. + \theta_{14} \theta_{23} \left[C^{a^\dagger a}(t_4,t_1) - C^{aa^\dagger}(t_1,t_4) \right]
\left[C^{a^\dagger a}(t_3,t_2) - C^{aa^\dagger}(t_2,t_3)\right]\right\} ,
\end{eqnarray}

\begin{eqnarray}
G^{2c}_{qccc}(t_1,t_2,t_3,t_4) & = & \frac{i}{2} \left\{ 
\theta_{12}\theta_{23}\theta_{34} \left(B^{aaa^\dagger a^\dagger}(t_1,t_2,t_3,t_4)
- B^{a^\dagger a^\dagger aa}(t_4,t_3,t_2,t_1) \right)\right. \nonumber \\
& & + \theta_{12}\theta_{24}\theta_{43} \left(B^{aaa^\dagger a^\dagger}(t_1,t_2,t_4,t_3)
- B^{a^\dagger a^\dagger aa}(t_3,t_4,t_2,t_1) \right) \nonumber \\
& & + \theta_{13}\theta_{32}\theta_{24} \left(B^{aa^\dagger aa^\dagger}(t_1,t_3,t_2,t_4)
- B^{a^\dagger aa^\dagger a}(t_4,t_2,t_3,t_1) \right) \nonumber \\
& & + \theta_{13}\theta_{34}\theta_{42} \left(B^{aa^\dagger a^\dagger a}(t_1,t_3,t_4,t_2)
- B^{aa^\dagger a^\dagger a}(t_2,t_4,t_3,t_1) \right) \nonumber \\
& & + \theta_{14}\theta_{42}\theta_{23} \left(B^{aa^\dagger aa^\dagger}(t_1,t_4,t_2,t_3)
- B^{a^\dagger aa^\dagger a}(t_3,t_2,t_4,t_1) \right) \nonumber \\
& & + \theta_{14}\theta_{43}\theta_{32} \left(B^{aa^\dagger a^\dagger a}(t_1,t_4,t_3,t_2)
- B^{aa^\dagger a^\dagger a}(t_2,t_3,t_4,t_1) \right) \nonumber \\
& & -\theta_{12}\theta_{13}\theta_{34} \left(B^{aaa^\dagger a^\dagger}(t_2,t_1,t_3,t_4)
- B^{a^\dagger a^\dagger aa}(t_4,t_3,t_1,t_2) \right) \nonumber \\
& & -\theta_{12}\theta_{14}\theta_{43} \left(B^{aaa^\dagger a^\dagger}(t_2,t_1,t_4,t_3)
- B^{a^\dagger a^\dagger aa}(t_3,t_4,t_1,t_2) \right) \nonumber \\
& & -\theta_{14}\theta_{13}\theta_{32} \left(B^{aa^\dagger aa^\dagger}(t_2,t_3,t_1,t_4)
- B^{a^\dagger aa^\dagger a}(t_4,t_1,t_3,t_2) \right) \nonumber \\
& & +\theta_{13}\theta_{14}\theta_{42} \left(B^{aa^\dagger aa^\dagger}(t_2,t_4,t_1,t_3)
- B^{a^\dagger aa^\dagger a}(t_3,t_4,t_1,t_2) \right) \nonumber \\
& & -\theta_{13}\theta_{12}\theta_{24} \left(B^{a^\dagger aaa^\dagger}(t_3,t_1,t_2,t_4)
- B^{a^\dagger aaa^\dagger}(t_4,t_2,t_1,t_3) \right) \nonumber \\
& & \left. +\theta_{14}\theta_{12}\theta_{23} \left(B^{a^\dagger aaa^\dagger}(t_3,t_2,t_1,t_4)
- B^{a^\dagger aaa^\dagger}(t_4,t_1,t_2,t_3) \right) \right\} .
\end{eqnarray}

\end{widetext}

\end{appendix}

\end{document}